\newif\ifarxivmode
\arxivmodetrue

\documentclass[reprint,aps,nofootinbib,superscriptaddress,floatfix,a4paper]{revtex4-1}

\usepackage{stmaryrd}
\usepackage{amsfonts,amsmath}
\usepackage[T1]{fontenc}

\usepackage[utf8]{inputenc}
\usepackage{graphicx,prettyref,subfigure}
\usepackage[separate-uncertainty=true]{siunitx}
\usepackage{tabularx,color,colortbl} %tables
\usepackage{xr}
\usepackage{chngcntr}

\usepackage{enumitem} %compact lists
\setlist{noitemsep,topsep=0pt,parsep=0pt,partopsep=0pt}

\usepackage{setspace}
\usepackage{todonotes}

\newcommand{\smalltodo}[2][]
{\todo[size=\footnotesize, caption={#2}, #1]{\begin{spacing}{0.5}#2\end{spacing}}}
\newcommand{\smalltodored}[2][]
{\smalltodo[color=red, #1]{#2}}
\newcommand{\smalltodoyellow}[2][]
{\smalltodo[color=yellow, #1]{#2}}
\newcommand{\smalltodogreen}[2][]
{\smalltodo[color=green, #1]{#2}}
\newcommand{\verysmalltodo}[2][]
{\smalltodo[size=\tiny, #1]{#2}}

\renewcommand{\thesubfigure}{(\Alph{subfigure})}

\newlength{\columnfigwidth}
\newlength{\fullfigwidth}
\newlength{\columnwidthleft}
\newlength{\columnwidthmiddle}

\newcommand{\modelhdr}[3]{
  \multicolumn{#1}{|l|}{
    \color{white}
    \cellcolor[gray]{0.0}
    \textbf{\makebox[0pt][l]{#2}\hspace{0.5\fullfigwidth}\makebox[0pt][c]{#3}}
  }
}
\newcommand{\parameterhdr}[3]{
  \multicolumn{#1}{|l|}{
    \color{black}\cellcolor[gray]{0.8}
    \textbf{\makebox[0pt][l]{#2}\hspace{0.5\fullfigwidth}\makebox[0pt][c]{#3}}
  }
}
\renewcommand{\arraystretch}{1.5}

\newcommand{\mytitle}{The effect of heterogeneity on decorrelation mechanisms in spiking neural networks:\\ a neuromorphic-hardware study}
\newcommand{\mytitlenobreak}{The effect of heterogeneity on decorrelation mechanisms in spiking neural networks: a neuromorphic-hardware study}
\newcommand{\myauthor}{Pfeil et al.}

\usepackage[colorlinks=true,linkcolor=black,citecolor=black,urlcolor=black,filecolor=black,
pdftitle={\mytitlenobreak{}},pdfauthor={\myauthor{}},pdfsubject={},pdfkeywords={},
bookmarks=false,pdffitwindow]{hyperref}

\newrefformat{cap}{\hyperref[#1]{Figure~\ref{#1}}}
\newrefformat{fig}{\hyperref[#1]{Figure~\ref{#1}}}
\newrefformat{sec}{\hyperref[#1]{Section~\ref{#1}}}
\newrefformat{sub}{\hyperref[#1]{Section~\ref{#1}}}
\newrefformat{tab}{\hyperref[#1]{Table~\ref{#1}}}
\newrefformat{eq}{\hyperref[#1]{Equation~\ref{#1}}}
\newrefformat{app}{\hyperref[#1]{\ref{#1}}}
\newrefformat{appfig}{\hyperref[#1]{Supplements Figure~\ref{#1}}}
\newrefformat{apptab}{\hyperref[#1]{Supplements Table~\ref{#1}}}

\newcommand{\spikey}{\emph{Spikey}}                      % Spikey
\newcommand{\pynn}{\texttt{PyNN}}                        % PyNN
\newcommand{\fb}{FB}                                     % feedback network
\newcommand{\fbrep}{$\text{FB}_{\text{replay}}$}         % replayed network
\newcommand{\rand}{RAND}                                 % replayed network with input randomized
\newcommand{\mrmsub}[2]{#1_{\text{#2}}}                  % non-italic subscripts
\newcommand{\sub}[2]{#1_{#2}}                            % italic subscripts
\newcommand{\mrmsup}[2]{#1^{\text{#2}}}                  % non-italic superscripts

\newcommand{\ti}{t^*}                                    % indexed spike
\newcommand{\runtime}{T}                                 % runtime (biol. domain)
\newcommand{\exectime}{\mrmsub{\runtime}{exe}}           % time needed for execution of emulation (wall clock domain)
\renewcommand{\vec}[1]{{\bf #1}}
\newcommand{\mat}[1]{{\bf #1}}
\newcommand{\cvJ}{CV_J}                                  % coefficient of variation of synaptic weights

\newcommand{\netsize}{N}                                 % network size
\newcommand{\netsizeE}{N_\text{E}}                       % size of excitatory population
\newcommand{\netsizeI}{N_\text{I}}                       % size of inhibitory population
\newcommand{\noinputs}{K}                                % in-degree of each neuron
\newcommand{\noinputsE}{K_\text{E}}                      % excitatory in-degree of each neuron
\newcommand{\noinputsI}{K_\text{I}}                      % inhibitory in-degree of each neuron
\newcommand{\neuronindex}{i}                             % indexed neuron
\newcommand{\spikesource}{\xi}                           % off-chip spike source
\newcommand{\spikesourceindex}{\sub{\spikesource}{i}}    % indexed spike source

\newcommand{\cmem}{\mrmsub{C}{m}}                        % membrane capacitance
\newcommand{\vm}{v}                                      % membrane potential
\newcommand{\gleak}{\mrmsub{g}{l}}                       % leakage conductance
\newcommand{\gleaknull}{\mrmsub{g}{l,0}}                 % leakage conductance (uncalibrated)
\newcommand{\vrest}{\mrmsub{E}{l}}                       % resting potential
\newcommand{\vthresh}{\Theta}                            % firing threshold
\newcommand{\vreset}{\mrmsub{\vm}{reset}}                % reset potential
\newcommand{\vinh}{\mrmsub{E}{inh}}                      % inhibitory reversal potential
\newcommand{\tauref}{\mrmsub{\tau}{ref}}                 % refractory time
\newcommand{\taumem}{\mrmsub{\tau}{m}}                   % membrane time constant
\newcommand{\taumemeff}{\mrmsup{\taumem}{\text{eff}}}    % effective membrane time constant

\newcommand{\ginh}{\mrmsub{g}{syn}}                      % synaptic conductance
\newcommand{\gmax}{\mrmsub{g}{max}}                      % max conductance of hardware synapse
\newcommand{\tausyn}{\mrmsub{\tau}{syn}}                 % synaptic time constant
\newcommand{\tausyneff}{\mrmsup{\tausyn}{\text{eff}}}    % effective synaptic-current time constant
\newcommand{\weighthw}{\mrmsub{w}{hw}}                   % synaptic weight (in hardware stored as digital value)
\newcommand{\weight}{J}                                  % synaptic weight
\newcommand{\weightE}{J_\text{E}}                        % synaptic weight (exc)
\newcommand{\weightI}{J_\text{I}}                        % synaptic weight (inh)
\newcommand{\delay}{d}                                   % synaptic delay
\newcommand{\delayeff}{\mrmsup{d}{\text{eff}}}           % effective synaptic delay
\newcommand{\effweight}{w}                               % effective weight
\newcommand{\vmaxeff}{\mrmsup{\mrmsub{V}{max}}{\text{eff}}} % maximum effective post-synaptic potential

\newcommand{\rateindex}{\sub{r}{i}}                  % time average of firing rate for one neuron
\newcommand{\ratetime}{$\rate$(t)}                       % population average of firing rate for time step

\newcommand{\avrateneuroncalib}{\meanrate}               % target firing rate
\newcommand{\meanmem}{\bar{\vm}}                         % time average of membrane potential
\newcommand{\stdmem}{\sigma(\vm)}                        % standard deviation of above
\newcommand{\calibfactor}{a}                             % heterogeneity
\newcommand{\compfactor}{b}                              % calibration factor
\newcommand{\cvisi}{\mrmsub{CV}{ISI}}                 % coefficient of variation of interspike intervals
\newcommand{\dist}{D}                                    % distance of average membrane potential to threshold in standard deviations of membrane potential
\newcommand{\ratewidth}{\Delta r}                        % width of firing rate distribution

\newcommand{\spiketrain}{s}                              % spike train
\newcommand{\rate}{r}                                    % firing rate of neuron
\newcommand{\popactivity}{\bar{\spiketrain}}             % population activity
\newcommand{\meanrate}{\bar{\rate}}                      % time-averaged population activity
\newcommand{\spiketrainfourier}{\MakeUppercase{\spiketrain}} % Fourier transform of spike train
\newcommand{\spiketrainindex}{\sub{\spiketrain}{i}}      % spike train with index
\newcommand{\memindex}{\sub{\vm}{i}}                     % spike train with index
\newcommand{\memfourier}{\MakeUppercase{\vm}}            % Fourier transform of membrane potential
\newcommand{\powerspec}{A}                               % population-averaged power spectrum
\newcommand{\poppowerspec}{\bar{A}}                      % population power spectrum
\newcommand{\crossspec}{C}                               % population-averaged cross spectrum
\newcommand{\crossfunction}{\MakeLowercase{\crossspec}}  % population-averaged cross-correlation function
\newcommand{\lfc}{\kappa}                                % low-frequency coherence
\newcommand{\fmin}{\mrmsub{f}{min}}                      % integral boundries for LFC
\newcommand{\fmax}{\mrmsub{f}{max}}
\newcommand{\connectivity}{\epsilon}                     % connectivity
\newcommand{\tbin}{\Delta t}                             % bin size spike train
\newcommand{\tbinmem}{\Delta t_\mathrm{m}}               % bin size membrane potential
\newcommand{\Twarmup}{T_\text{warmup}}                   % burn-in time
\newcommand{\Mtrials}{M}                                 % number of network realizations
\newcommand{\Ltrials}{L}                                 % number of trials
\newcommand{\Df}{\Delta F}                               % sliding window to smoothen coherence

\newcommand{\taumax}{\tau_\mathrm{max}}                  % integration window for effective weights after trigger spikes
\newcommand{\taumin}{\tau_\mathrm{min}}                  % integration window for effective weights before trigger spikes

\renewcommand{\r}{\vec{r}}
\newcommand{\W}{\mat{W}}
\newcommand{\x}{\vec{x}}
\newcommand{\h}{h}
\newcommand{\diag}{\mathrm{diag}}
\newcommand{\CRR}{\mat{C}_\mathrm{RR}}
\newcommand{\CRRIN}{\mat{C}_\mathrm{RR}^\mathrm{in}}
\newcommand{\rff}{\tilde{\r}}
\newcommand{\q}{\vec{q}}
\newcommand{\CRRFF}{\mat{C}_\mathrm{\tilde{R}\tilde{R}}}
\newcommand{\CQQIN}{\mat{C}_\mathrm{QQ}^\mathrm{in}}
\newcommand{\CQQ}{\mat{C}_\mathrm{QQ}}
\newcommand{\ABARX}{\bar{A}_\mathrm{X}}
\newcommand{\CXXii}{C_{\mathrm{XX},ii}}
\newcommand{\CBARXX}{\bar{C}_\mathrm{XX}}
\newcommand{\CXXij}{C_{\mathrm{XX},ij}}
\newcommand{\muw}{\mu_\mathrm{\effweight}}
\newcommand{\sigmaw}{\sigma_\mathrm{\effweight}}
\newcommand{\muW}{\mu_\mathrm{\MakeUppercase{\effweight}}}
\newcommand{\sigmaW}{\sigma_\mathrm{\MakeUppercase{\effweight}}}

\newcommand{\dither}{\vartheta}
\newcommand{\stdweight}{\sigma_\weight}
\newcommand{\stdthresh}{\sigma_\vthresh}
\newcommand{\aveffweight}{\bar{\effweight}}

\newcommand{\Isyn}{\mrmsub{I}{syn}}
\newcommand{\numrecordspikesE}{P_\mathrm{\spiketrain}^\text{E}}
\newcommand{\numrecordmemE}{P_\mathrm{\vm}^\text{E}}
\newcommand{\numrecordspikesI}{P_\mathrm{\spiketrain}^\text{I}}
\newcommand{\numrecordmemI}{P_\mathrm{\vm}^\text{I}}
\newcommand{\numrecordspikes}{P_\mathrm{\spiketrain}}
\newcommand{\numrecordmem}{P_\mathrm{\vm}}

\definecolor{graytarget}{gray}{0.5}
\ifarxivmode
\else
\fi

\begin{document}

\begin{abstract}

High-level brain function such as memory, classification or reasoning can be realized by means of recurrent networks of simplified model neurons.
Analog neuromorphic hardware constitutes a fast and energy efficient substrate for the implementation of such neural computing architectures in technical applications and neuroscientific research.
The functional performance of neural networks is often critically dependent on the level of correlations in the neural activity.
In finite networks, correlations are typically inevitable due to shared presynaptic input.
Recent theoretical studies have shown that inhibitory feedback, abundant in biological neural networks, can actively suppress these shared-input correlations and thereby enable neurons to fire nearly independently.
For networks of spiking neurons, the decorrelating effect of inhibitory feedback has so far been explicitly demonstrated only for homogeneous networks of neurons with linear sub-threshold dynamics.
Theory, however, suggests that the effect is a general phenomenon, present in any system with sufficient inhibitory feedback, irrespective of the details of the network structure or the neuronal and synaptic properties.
Here, we investigate the effect of network heterogeneity on correlations in sparse, random networks of inhibitory neurons with non-linear, conductance-based synapses.
Emulations of these networks on the analog neuromorphic hardware system \spikey{} allow us to test the efficiency of decorrelation by inhibitory feedback in the presence of hardware-specific heterogeneities.
The configurability of the hardware substrate enables us to modulate the extent of heterogeneity in a systematic manner.
We selectively study the effects of shared input and recurrent connections on correlations in membrane potentials and spike trains.
Our results confirm that shared-input correlations are actively suppressed by inhibitory feedback also in highly heterogeneous networks exhibiting broad, heavy-tailed firing-rate distributions.
In line with former studies, cell heterogeneities reduce shared-input correlations. 
Overall, however, correlations in the recurrent system can increase with the level of heterogeneity as a consequence of diminished effective negative feedback.

\end{abstract}
\newcommand{\affkip}{Kirchhoff-Institute for Physics,
                     Heidelberg University,
                     Heidelberg, Germany}
\newcommand{\afffzj}{Institute of Neuroscience and Medicine (INM-6) and 
                               Institute for Advanced Simulation (IAS-6) and 
                               JARA BRAIN Institute I, 
                               J\"ulich Research Centre,    
                               J\"ulich, Germany}
\newcommand{\affrwthmed}{Department of Psychiatry, 
                                    Psychotherapy and Psychosomatics, 
                                    Medical Faculty, 
                                    RWTH Aachen University, 
                                    Aachen, Germany}
\newcommand{\affrwthphys}{Department of Physics, 
                                     Faculty 1, 
                                     RWTH Aachen University, 
                                     Aachen, Germany }

\newcommand{\sharedauthor}{These authors contributed equally to this study.}

\renewcommand{\thefootnote}{\fnsymbol{footnote}}
\newcounter{superscriptcounter}
\setcounter{superscriptcounter}{2}

\title{\mytitle}

\author{Thomas Pfeil\textsuperscript{\fnsymbol{superscriptcounter}}}
\thanks{\sharedauthor}
\affiliation{\affkip}
\author{Jakob Jordan}
\thanks{\sharedauthor}
\affiliation{\afffzj}
\author{Tom Tetzlaff}
\affiliation{\afffzj}
\author{Andreas Gr\"ubl}
\affiliation{\affkip}
\author{Johannes Schemmel}
\affiliation{\affkip}
\author{Markus Diesmann}
\affiliation{\afffzj}
\affiliation{\affrwthmed}
\affiliation{\affrwthphys}
\author{Karlheinz Meier}
\affiliation{\affkip}

\date{\today}

\maketitle

\footnotetext[\value{superscriptcounter}]{
    Correspondence:\hspace{1em}\parbox[t]{10cm}{
    Thomas Pfeil\\
    Kirchhoff-Institute for Physics\\
    Im Neuenheimer Feld 227\\
    69120 Heidelberg, Germany\\
    tel: +49-6221-549813\\
    \href{mailto:thomas.pfeil@kip.uni-heidelberg.de}{thomas.pfeil@kip.uni-heidelberg.de}}
}

\renewcommand{\thefootnote}{\number\value{footnote}}

\section{Introduction}

Dynamical systems in nature often exhibit a remarkable degree of diversity, specialization or anticorrelation across their components,
despite equalizing factors such as common input or homogeneity in component and interaction parameters.
In many cases, these observations can be explained by the effect of negative feedback.
Cell differentiation caused by lateral inhibition \citep{Gierer1974}, formation of new species driven by competition \citep{Rosenzweig1978} or antiferromagnetism \citep{Neel1952} constitute just a few examples.
In recurrent neuronal networks, inhibitory feedback constitutes a powerful decorrelation mechanism which allows different neurons to respond nearly independently, even if they
are driven by largely overlapping local or external inputs \citep{Renart10_587,Tetzlaff12_e1002596,Helias14}.
Decorrelation by negative feedback hence implements an efficient form of redundancy reduction.
In biological systems, it may serve similar purposes as decorrelation in technical applications, where it is used in data compression (e.g., principal-component analysis \citep{Pearson1901}), crosstalk reduction (e.g., in digital signal processing \citep{Ginis06_016828}), echo suppression (e.g., in acoustics \citep{Benesty01}) or random-number generation in hardware \citep{Trichina01_26}.
Moreover, inhibitory feedback suppresses ``quantization noise'' at low frequencies and can thereby increase the dynamical range and signal-to-noise ratio for the encoding of analog signals in the spiking activity of recurrent neural networks \citep{Mar99_10450}.
It is tempting to exploit these mechanisms in synthetic, neurally inspired architectures such as analog neuromorphic hardware.

Analog neuromorphic hardware mimics properties of biological neural systems using physical models of neurons and synapses (capacitors, for example, emulate insulating cell membranes) \citep{Mead90_1629,Indiveri11_00073}.
The temporal evolution of the analog circuits represents a solution to the corresponding model equations.
In consequence, neural-network emulations on analog neuromorphic hardware are massively parallel, extremely fast and energy efficient.
Analog neuromorphic devices are therefore highly attractive as tools for neuroscientific research, e.g., for the investigation of learning on long time scales, and technical applications \citep{Sensemaker05,FACETS10,Brainscales14,HumanBrainProject14}.
A biologically inspired neural network (olfactory system of insects) performing rapid online data (odor) classification, for example, has recently been successfully implemented on the analog neuromorphic hardware system \spikey{} \citep{Pfeil13_11,Schmuker14_2081}.
In this application, decorrelation by inhibition is an essential ingredient to guarantee high classification performance.
The suppression of quantization noise by inhibitory feedback \citep{Mar99_10450} has been used as a means of noise shaping in several neuromorphic-hardware applications, aiming at the construction of biologically inspired ultra-low power analog-to-digital converters \citep{Marienborg2002_1010643,Kikombo2009_5178715}.

For the functional performance of neuronal architectures, the level of correlations between the activities of individual neurons is often pivotal.
Whether such correlations are beneficial or not is context dependent.
A number of previous studies emphasize a functional benefit of certain types of correlation for encoding/decoding of information in/from populations of neurons \citep{Ecker11_14272,Averbeck06_358,Morenobote14_1410}, information transmission \citep{Fries05_474,Abeles91,Diesmann99_529}, robustness against noise \citep{Tabareau10_1000637}, or gain control of postsynaptic neurons \citep{Salinas01}.
Other studies argue that positive cross-correlations are detrimental as they decrease the precision or sparseness of population codes \citep{Shadlen98,Abbott99,Tripp07_1830,Schmuker07_20285,Schmuker14_2081}.
\citet{Cohen09_1079}, for example, have shown that decreased spike-train correlations in macaque visual area V4 are accompanied by increased behavioral performance in an orientation change-detection task.
Depending on the similarity between the trial-averaged responses of different neurons to external stimuli (signal correlation), noise correlations (correlations not explained by signal correlations) can either increase or decrease the amount of information that can be encoded in or decoded from a population of neurons.
In populations of neurons with high signal correlation, vanishing or even negative noise correlations are desirable to improve the population code \citep{Averbeck06_358}.

In finite neural networks, an inevitable source of correlated neural activity is common presynaptic input, shared by multiple postsynaptic neurons.
In network models and in-vivo recordings, however, pairwise correlations in the activity of neighboring neurons have been found to be substantially smaller than expected given the amount of shared input \citep{Tetzlaff04_117,Kriener08_2185,Tetzlaff08_2133,Ecker10,Renart10_587,Tetzlaff12_e1002596}.
In several studies, this observation has been explained by inhibitory coupling.
While \citet{Ly_12} and \citet{Middleton_12} primarily focused on the effect of feedforward inhibition,
\citet{Renart10_587}, \citet{wiechert2010_1003}, and \citet{Tetzlaff12_e1002596} attributed the smallness of correlations to an active decorrelation of neural activity by inhibitory feedback.
The mechanism underlying this active decorrelation has already been described by \citet{Mar99_10450}.
In this study, the authors focused on the suppression of low-frequency fluctuations of the population firing rate by recurrent dynamics.
As the amplitude of population-rate fluctuations is directly linked to pair-wise correlations (see, e.g., \citep{Harris11_509}), the effect described in \citep{Mar99_10450} corresponds to a suppression of pairwise correlations in the spiking activity.
The theory underlying decorrelation by inhibitory feedback suggests the effect to be general:
Decorrelation should be observable in any system with sufficiently strong inhibitory feedback, irrespective of the details of the network structure and the cell and synapse properties.
For networks of spiking neurons, however, the effect has so far been explicitly demonstrated only for the homogeneous case, where all neurons have identical properties, receive (approximately) the same number of inputs, and, hence, fire at about the same rate \citep{Renart10_587,Tetzlaff12_e1002596}.
Moreover, the sub-threshold dynamics of individual neurons was assumed to be linear.

Biological neuronal networks typically exhibit broad, heavy-tailed firing-rate distributions \citep{Griffith1966_516,Koch1989_292,Shafi2007_1082,Hromadka2008_0060016,OConnor2010_1048,Roxin11_16217,buzsaki14_264}, indicating a high degree of heterogeneity, e.g., in synaptic weights \citep{Song05_0507,Lefort09_301,Koulakov09_3685,Avermann12,Ikegaya13_293}, in-degrees \citep{Roxin11_8} or time constants \citep{Kuhn04, Roxin11_16217}.
The same holds for neural networks implemented on analog neuromorphic hardware.
All analog circuits suffer from device variations caused by unavoidable variability in the manufacturing process.
Neurons and synapses implemented in analog neuromorphic hardware therefore exhibit heterogeneous response properties, similar to their biological counterparts \citep{Stein05,Marder06_563}.
To understand the dynamics and function of recurrent neural networks in both biological and synthetic substrates, it is therefore essential to account for such heterogeneities.

Previous work on recurrent neural networks has shown that heterogeneity in single-neuron properties or connectivity broadens the distribution of firing rates \citep{VanVreeswijk98_1321,Roxin11_16217} and affects the stability of asynchronous or oscillatory states \citep{Tsodyks1993_1280,Golomb93_4810,neltner2000_1607,Denker04,Roxin11_8,Mejias2012_228102}.
A number of studies pointed at a potential benefit of heterogeneity for the information-processing capabilities of neural networks \cite{Stocks2000,Shamir2006,Chelaru2008_16344,Osborne2008,Padmanabhan10_1276,Marsat2010,Holmstrom2010,Yim2013_032710,Mejias2012_228102,lengler2013_e80694,Mejias2014,Tripathy14052013}.
The effect of heterogeneity on correlations in the activity of recurrent networks of spiking neurons, however, remains unclear.
\citet{Padmanabhan10_1276} have shown that the responses of a population of unconnected neurons are decorrelated by heterogeneity in the neuronal response properties.
These results are supported by the subsequent theoretical analysis in \citep{Yim2013_032710}.
In the following, we refer to this type of decorrelation by heterogeneity as \emph{feedforward decorrelation}.
It does not account for the effect of the recurrent network dynamics.
Active decorrelation due to inhibitory feedback \citep[see above;][]{Renart10_587, Tetzlaff12_e1002596}, in contrast, constitutes a very different mechanism.
The effect of heterogeneity on this \emph{feedback decorrelation} has lately been studied by \citet{Bernacchia13_1732} in the framework of a recurrent network of linear firing-rate neurons.
In this setup, correlations are suppressed by heterogeneity in the network connectivity (distributions of coupling strengths or random dilution of connectivity).
It remains unclear, however, whether this holds true for networks of (nonlinear) spiking neurons.

\begin{figure}
\includegraphics[width=\columnfigwidth]{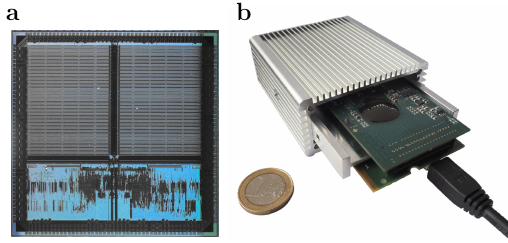}
\caption{
The neuromorphic hardware system \spikey{}.
\textbf{(a)}
Photograph of the \spikey{} chip (size $\SI{5}{} \times \SI{5}{\milli\meter\squared}$).
It comprises analog circuits of $384$ neurons and $98304$ synapses,
is highly configurable and emulates neural-network dynamics with a speed-up of $10^4$ with respect to biological real-time.
\textbf{(b)}
Photograph of the partly cased \spikey{} system, carrying the \spikey{} chip (covered by a black round seal) and conventional memory. %composed by two printed circuit boards.
The system is connected to the host computer via USB 2.0, consumes \SI{6}{\watt} of power in total and less than \SI{1}{\nano\joule} per synaptic transmission (see \prettyref{app:power}).
}
\label{fig:hardware}
\end{figure}

In this study, we investigate the impact of heterogeneity on input and output correlations in the asynchronous regime of sparse networks of leaky integrate-and-fire (LIF) neurons with conductance-based synapses.
Emulation of the networks on the analog neuromorphic hardware system \spikey{} (\prettyref{fig:hardware}) \citep{Schemmel06_1,Pfeil13_11} enable us to investigate the impact of substrate specific properties on the network dynamics.
Insights about the interplay between features of the computing substrate and network dynamics are a necessary prerequisite for the development of algorithms that exploit the benefits of analog neuromorphic systems at best.
The configurability of this system \citep{Pfeil13_11} enables us to systematically vary the level of heterogeneity, and to disentangle the effects of heterogeneity on feedforward and feedback decorrelation (see above).
For simplicity, we focus on purely inhibitory networks, thereby emphasizing that active decorrelation by inhibitory feedback does not rely on a dynamical balance between excitation and inhibition \citep{Tetzlaff12_e1002596,Helias14}.
We show that decorrelation by inhibitory feedback is effective even in highly heterogeneous networks with broad distributions of firing rates (\prettyref{sub:results_decorrelation}).
Increasing the level of heterogeneity has two effects: Feedforward decorrelation is enhanced, feedback decorrelation is impaired.
Due to the latter, the overall input and output correlations do not necessarily become smaller with increasing heterogeneity. They can even increase (\prettyref{sub:results_heterogeneity}).

Note that results from specific network emulations on hardware do not directly translate to those obtained by simulations on conventional computers, because the dynamics, parametrization and interplay of analog circuits is very complex and difficult to reproduce with classical simulations.
If simplified models for spatial and temporal variability are considered in software simulations, however, emulation results can be reproduced qualitatively, thereby verifying the design of the hardware system.
While our hardware system is designed to physically implement biologically realistic neural algorithms in a fast and energy-efficient way,
software simulations are used as a complementary tool to isolate, verify and investigate different hardware features, such as spatial and temporal parameter variations.
Due to the limited access and configurability of network parameters, this would be difficult to achieve with hardware studies alone.
In analogy to the necessity of performing experiments on biological neural systems to verify assumptions made in Computational Neuroscience, actual emulations on neuromorphic hardware are essential to understand its properties and develop efficient neural algorithms for these devices.
The fact that our main findings hold true for both emulations on hardware and simulations with software, and that they can be distilled to simple linear models, support their broad relevance and robustness.

\section{Methods}
\label{sec:methods}

\subsection{Network model}
\label{sub:methods_network}

Details on the network, neuron and synapse model are provided in \prettyref{tab:nordlie}. 
Parameter values are given in \prettyref{tab:nordlie_params}.
Briefly: We consider a purely inhibitory, sparse network of $\netsize$ ($\netsize=192$, unless stated otherwise) LIF neurons with conductance-based synapses. 
Each neuron receives input from a fixed number $\noinputs=15$ of randomly chosen presynaptic sources, independently of the network size $\netsize$.
Self-connections and multiple connections between neurons are excluded.
Resting potentials $\vrest$ are set above the firing thresholds $\vthresh$ (equivalent to applying a constant supra-threshold input current).
We thereby ensure autonomous firing in the absence of any further external input.
Due to temporal noise, the initial conditions are essentially random.

\subsection{Network emulations on the neuromorphic-hardware system \spikey{}}
\label{sub:methods_hardware}

The \spikey{} chip (\prettyref{fig:hardware}) consists of physical models of LIF neurons and conductance-based synapses with exponentially decaying dynamics (for details, see \prettyref{tab:nordlie}).
The emergent dynamics of these physical models represents a solution for the model equations of neurons and synapses in continuous time, in parallel for all units.
In contrast, in classical simulations on von-Neumann architectures, model equations are solved by step-wise numerical integration, where parallelization is limited by the available number of processor cores.
To emphasize the difference between simulations using software and simulations using physical models, the term \emph{emulation} is used for the latter \citep{Pfeil13_11}.

The response properties of physical neurons and synapses vary across the chip due to unavoidable variations in the production process
that manifest in a spatially disordered pattern (\emph{fixed-pattern noise}).
In contrast to the approximately static fixed-pattern noise, \emph{temporal noise}, including electronic noise and transient experiment conditions (e.g., chip temperature), impairs the reproducibility of emulations.
In general, two network emulations with identical configuration and stimulation do not result in identical network activity.
Both fixed-pattern and temporal noise need to be taken into account when developing models for analog neuromorphic hardware.

The key features of the \spikey{} chip are the high acceleration and configurability of the analog network implementation.
Some network parameters, e.g., synaptic weights and leak conductances, are configurable for each unit, while other parameters are shared for several units (for details see \citep{Pfeil13_11}).
The hardware system is optimized for spike in- and output and allows to record the membrane potential of one (arbitrarily chosen) neuron with a sampling frequency of \SI{96}{\mega\hertz} in hardware time.
On the \spikey{} chip, capacitances are smaller and conductances are much higher than in biological nervous systems. In consequence, networks on the \spikey{} chip are emulated with a speed-up of approximately $10^4$ with respect to biological real-time.
Due to this high acceleration of the neuromorphic chip, the data bandwidth of the connection between the neuromorphic system and the host computer is not sufficient to communicate with the chip in real time.
Consequently, input and output spikes (for stimulation and from recordings, respectively) are buffered in a local memory next to the chip.
The high acceleration of the \spikey{} chip allows most of the transistors to operate outside of weak inversion, thereby reducing the effect of transistor variations and minimizing fixed-pattern noise.

In contrast to such accelerated systems, most other configurable, analog neuromorphic substrates are designed for real-time emulations at very low power consumption \citep{Badoni06_4,Hafliger07_551,Vogelstein07_253,Indiveri09_119,Serrano09_1417,Brink13_71,Benjamin14_699} and implement fewer, but more complex, neurons \citep{Renaud10_905,Yu10_139}.

Access to the \spikey{} system is encapsulated by the simulator-independent language \pynn{} \citep{Davison09, Bruederle09_17}, providing a stable and user-friendly interface.  
\pynn{} integrates the hardware into the computational neuroscience tool chain and has facilitated the implementation of several network models on the \spikey{} chip \citep{Kaplan09_1524, Bill10_129, Bruederle10_iscas, Pfeil13_11, Schmuker14_2081}.

\begin{figure*}
\includegraphics[width=\fullfigwidth]{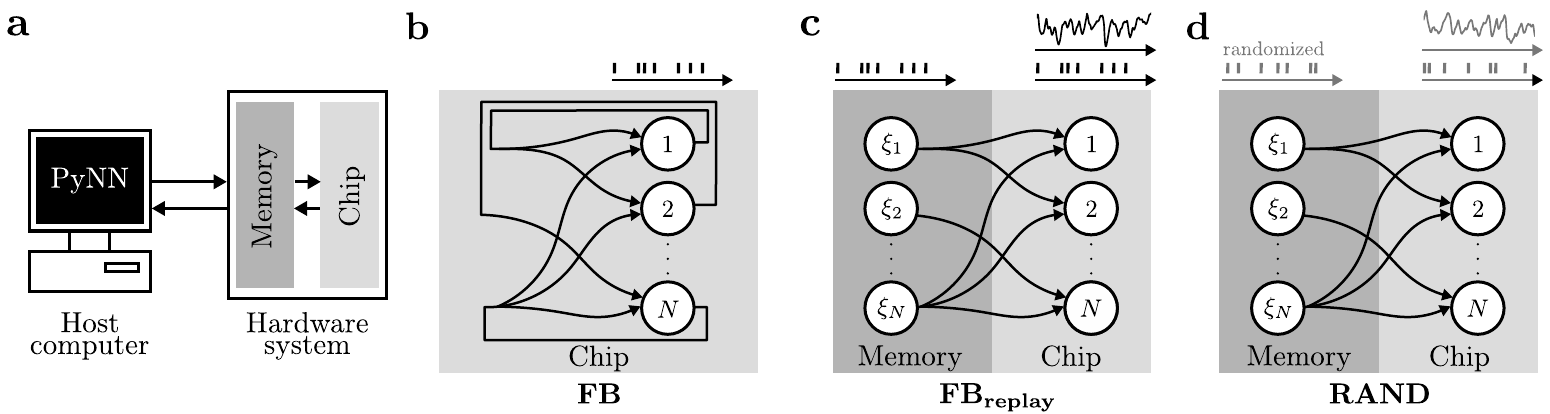}
\caption{
Experimental setup.
\textbf{(a)}
Data flow of the \spikey{} system.
For details see \prettyref{sub:methods_hardware}.
\textbf{(b)}
Network with on-chip feedback connections (\fb{}).
Spikes from all neurons are recorded to the local memory.
\textbf{(c)}
Spikes of the \fb{} network in (b) replayed from memory via off-chip spike sources $\spikesourceindex$ to neurons $\neuronindex$ (\fbrep{}).
Spike times of $\spikesourceindex$ correspond to those recorded from neuron $\neuronindex$ in (b).
Spikes from all neurons or the free membrane potential of one selected neuron are recorded.
\textbf{(d)}
Like (c), but spike times from (b) are randomized for each source (\rand{}).
}
\label{fig:topo}
\end{figure*}

On the \spikey{} system, a spiking neural network is emulated as follows (\prettyref{fig:topo}a):
First, the network described in \pynn{} is mapped to the \spikey{} chip, i.e., neurons and synapses are allocated and parametrized.
Second, input spikes, if available, are prepared on the host computer and transferred to the local memory on the hardware system.
Third, the emulation is triggered and available input spikes are generated.
Output spikes and membrane data are recorded to local memory.
Last, spike and membrane data are transferred to the host computer and scaled back into the biological domain of the \pynn{} model description.

For consistency with the model description and simplified comparison to the existing literature,
all hardware times and all hardware voltages are expressed in terms of the quantities they represent in the neurobiological model, throughout this study.

\subsection{Experimental setup}

To differentiate and compare the effects of shared inputs and feedback connections on correlations, we investigate two different emulation scenarios:
First, we emulate networks with intact feedback (\fb{}, \prettyref{fig:topo}b),
and second, the contribution of shared input is isolated by \emph{randomizing} the temporal order of this feedback (\rand{}, \prettyref{fig:topo}d).

In the \rand{} scenario, the inputs of neurons are decoupled from their outputs.
Spatio-temporal correlations in presynaptic spike trains are removed by randomizing the presynaptic spike times.

Input correlations between neurons are measured via their \emph{free membrane potential}, i.e., the membrane potential with disabled spiking mechanism (technically, the threshold is set very high).
Because membrane potential traces can be recorded in the hardware only one at a time,
traces are obtained consecutively, while repeatedly \emph{replaying} the previously recorded activity of the \fb{} network to a population of unconnected neurons of equal size.
We keep the connectivity the same, and hence each neuron receives the same number of spikes as in the recurrent network during the whole emulation,
either without (\fbrep{}, \prettyref{fig:topo}c) or with randomization of presynaptic spike times (\rand{}, \prettyref{fig:topo}d), respectively.
To preserve the fixed pattern of variability of synaptic weights in hardware, the same hardware synapses are used for each connection in both scenarios.
If network dynamics were reproduced perfectly, membrane potential traces and spike times would be identical in the \fb{} and \fbrep{} cases (see also \prettyref{sub:methods_reprod}).

Drawing two different network realizations (i.e., the connectivity matrix) results in the allocation of different hardware synapses,
and, due to fixed-pattern noise, in different values of synaptic weights.
To average over this variability, throughout this study, emulation results are averaged over $\Mtrials=100$ network realizations, if not stated otherwise.

\subsection{Reproducibility of hardware emulations}
\label{sub:methods_reprod}

Since the initial conditions of the recurrent network on hardware are undefined, consecutive emulations of the \fb{} network result in different network activities.
In the \rand{} and \fbrep{} case, however, the input of neurons is decoupled from their output.
Although unavoidable temporal noise is present, the system's state-space trajectory returns to the trajectory of the previously recorded \fb{} case.
A certain degree of reproducibility is required for two reasons:
First, the investigated effect of decorrelation by inhibitory feedback requires a precise relation between spike input and output.
Thus our method of replacing the feedback loop by replay is only valid if temporal noise does not substantially corrupt this relationship.
Second, to record the membrane potentials of all neurons, as if recorded at once, neuron dynamics have to be reasonably similar in consecutive emulations.

We measure the reproducibility of neuron dynamics by comparing
consecutive emulations with identical configuration, i.e., connectivity and stimulation.
For this purpose the spiking activity of a \fb{} network is first
recorded (\prettyref{fig:topo}b) and then repeatedly replayed
(\prettyref{fig:topo}c).
Reproducibility is quantified by the correlations ($\lfc_X$
in \prettyref{tab:anasummary}) of free membrane
potential traces and output spike trains obtained for individual neurons in $\Ltrials=25$ different trials.

\begin{figure}
\includegraphics{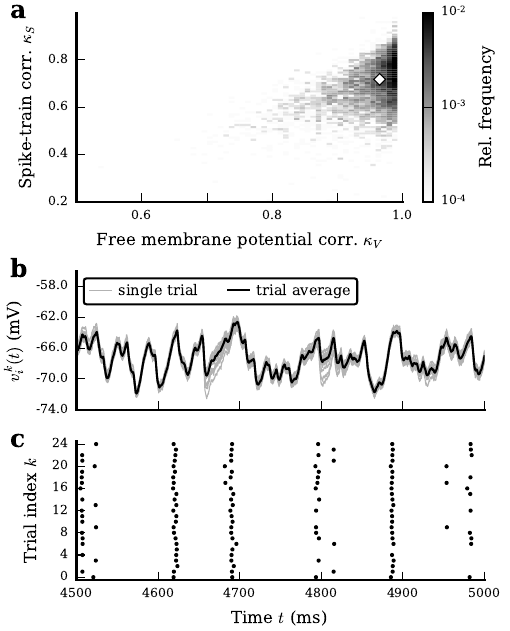}
\caption{
Reproducibility of free membrane potentials and spiking activity in the \fbrep{} case.
\textbf{(a)}
Low-frequency coherence $\lfc_{\memfourier}$ and $\lfc_{\spiketrainfourier}$ of free membrane potentials $\memindex^k(t)$ and $\memindex^l(t)$
and binned spike trains $\spiketrainindex^k(t)$ and $\spiketrainindex^l(t)$, respectively,
for each neuron $\neuronindex$ averaged over $\Ltrials=25$ trials $k,l$ with $k\neq l$,
for $\Mtrials=50$ different network realizations.
The diamond marks the average across all neurons $i$ and $\Mtrials$ network realizations ($\lfc_{\memfourier}=0.96$, $\lfc_{\spiketrainfourier}=0.72$).
\textbf{(b)}
Free single-trial membrane potentials $\memindex^k(t)$ (gray) and average over trials $\frac{1}{\Ltrials}\sum_{k=1}^{\Ltrials} \memindex^k(t)$ (black),
and \textbf{(c)} spike density $\spikesourceindex(t)$ of a single neuron $\neuronindex$ for $\Ltrials=25$ identical trials.
The selected neuron $\neuronindex$ has membrane potential coherence and spike train coherence closest to the diamond in (a).
}
\label{fig:reprod}
\end{figure}

Free membrane potentials are reproduced quite well, while spike trains show larger deviations across trials (\prettyref{fig:reprod}).
Small deviations in the membrane potential (\prettyref{fig:reprod}b) are amplified by the thresholding procedure \citep{Mainen95_1503,DeLaRocha07_802,wiechert2010_1003} and can lead to large differences between spike trains (\prettyref{fig:reprod}c).
Consequently, measures based on data of several consecutive replays are more precise for membrane potentials than for spike trains.
Nevertheless, results have to be interpreted with care in both cases.

\subsection{Calibration}
\label{sub:methods_calib}

The heterogeneity of the \spikey{} hardware is adjusted by calibrating the leak conductance\footnote{since capacitances and potentials can not be configured individually for each hardware neuron \citep{Pfeil13_11}} for each individual neuron, compensating for fixed-pattern noise of neuron parameters.
To this end, a population of unconnected neurons is driven by a constant supra-threshold current and the time-averaged population activity $\avrateneuroncalib$ is measured.
Then, we applied the bisection method \citep{Press07} to adjust the leak conductance $\gleak$ of each neuron, such that the neuron's firing rate matches the target rate $\avrateneuroncalib$.
This results in calibration values $\compfactor$ for the leak conductance $\gleak = \gleaknull(1 + \compfactor)$, where $\gleaknull$ is the leak conductance before calibration.
Because emulations on hardware are not perfectly reproducible,
more precise calibration was achieved by evaluating the median over $25$ identically configured trials instead of single trials.
Furthermore, the bisection method was modified for noisy systems (for details, see \prettyref{app:binary_noisy}).

Intermediate calibration states are obtained by linearly scaling the full calibration:
\begin{equation}
  \gleak = \gleaknull(1 + (1 - \calibfactor) \compfactor) \quad.
  \label{eq:calib}
\end{equation}
The heterogeneity $\calibfactor$ is chosen in $[0,1]$ for calibrations between the uncalibrated ($\calibfactor=1$) and calibrated state ($\calibfactor=0$).
In the following, the fully calibrated chip ($\calibfactor=0$) is used, if not stated otherwise.

\begin{figure}
\includegraphics{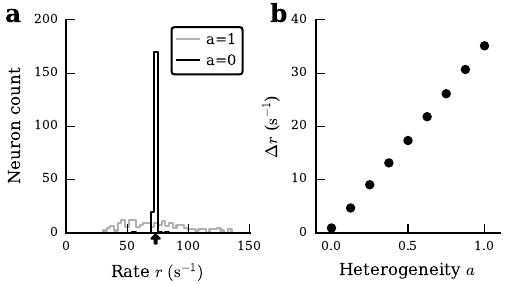}
\caption{
  Calibration of the \spikey{} chip.
  \textbf{(a)}
  Histogram of firing rates $\rate$ for a population of unconnected neurons with supra-threshold input currents, before (gray) and after (black) calibration, each neuron averaged over $\Ltrials=100$ trials.
  The arrow denotes the target rate $\avrateneuroncalib$.
  \textbf{(b)}
  Difference $\ratewidth = r_{P75} - r_{P25}$ of 75th and 25th percentile of the histograms in (a), as a function of network heterogeneity $\calibfactor$ (\prettyref{eq:calib}).
  The mean firing rate over all values of $\calibfactor$ is \SI{73.4 \pm 0.3}{\per\second}.
}
\label{fig:calib}
\end{figure}

This calibration substantially narrows the distribution of firing
rates compared to the uncalibrated state (\prettyref{fig:calib}).
With respect to the stationary firing rate, variability on the neuron level
is reduced from 35.1 to $\SI{0.9}{\second}^{-1}$.

Even in the fully calibrated state, leak conductances can still be widely distributed.
Due to the chosen calibration procedure, they are likely to be correlated to other parameters that influence the neurons' response to a constant supra-threshold current after calibration.
This mutual compensation can lead to similar phenomenology (here: firing rates) despite disparate parameter values, similar to what is observed in biology \citep{Prinz-2004_1345}.
In addition to remaining variations in neuron parameters, synaptic parameters are significantly distributed \citep{Pfeil13_1,Schmuker14_2081}.

\subsection{Correlation measures}
\label{sub:methods_decorr}

In the following, we introduce definitions used to analyze the recorded data.
For clarity, all relevant equations and their parametrization are listed in \prettyref{tab:anasummary} and \ref{tab:anaparams}, respectively.

We quantify correlations of membrane potentials $\memindex(t)$ and spike trains $\spiketrainindex(t)$ by the population-averaged \emph{low-frequency coherence} $\lfc_{\memfourier}$ and $\lfc_{\spiketrainfourier}$, respectively.
At frequency zero, the coherence corresponds to the normalized integral of the cross-covariance function, i.e., it measures correlations on all time scales.
We define the low-frequency coherence $\lfc_{X}$, with $X\in \{S,V\}$, to be the average coherence over a frequency interval from \SI{0.1}{} to \SI{20}{\hertz}.
In this interval, the suppression of population-rate fluctuations in recurrent networks due to inhibitory feedback is most pronounced, and the coherence is approximately constant.
Before calculating the coherence, we convolve the power- and cross-spectra with a rectangular window to average out random fluctuations.
This measure, or a variant of it, is commonly used in the neuroscientific literature \cite{Bair01,Kohn05,Morenobote06_028101,DeLaRocha07_802,Renart10_587,Tetzlaff12_e1002596,Yim2013_032710}.
We use the terms low-frequency coherence and correlation interchangeably.

Throughout this study, the term \emph{input correlations} is used for correlations between free membrane potentials,
and \emph{output correlations} for correlations between spike trains.
\emph{Shared-input correlations} are membrane-potential correlations that are exclusively caused by overlapping presynaptic sources, ignoring possible correlations in the presynaptic activity.
In homogeneous networks, the average pairwise shared-input correlation 
\begin{equation}
  \lfc_{\memfourier} = \frac{\noinputs}{\netsize}
  \label{eq:shared_input}
\end{equation}
is given by the connectivity $K/N$ \citep{Tetzlaff12_e1002596}.
In heterogeneous networks, shared-input correlations can be reduced.
In the presence of heterogeneous synaptic weights, for example, the shared-input correlation
\begin{equation}
  \lfc_{\memfourier} = \frac{1}{1 + \cvJ^2} \frac{\noinputs}{\netsize}
  \label{eq:shared_input_hetero}
\end{equation}
is decreased by a factor $1/(1 + \cvJ^2)$,
where $\cvJ$ denotes the coefficient of variation of the (non-zero) synaptic weights.
Note, however, that heterogeneities which affect only the spike generation but not the integration of synaptic inputs, e.g.~distributions of firing thresholds, have no effect on the shared-input correlation.

We assess the significance of correlations by comparing the results from emulations to correlations in surrogate data, in which we removed spatial correlations.
For every neuron, we randomly shuffled bins of the membrane potential trace, and assigned a new timestamp uniformly drawn from the emulation interval to every spike, respectively.
We thereby remove all spatio-temporal correlations between neurons recorded in parallel.
By this procedure we create $100$ surrogate trials, across which we calculate the average correlations and the standard error.

\begin{figure*}
\includegraphics{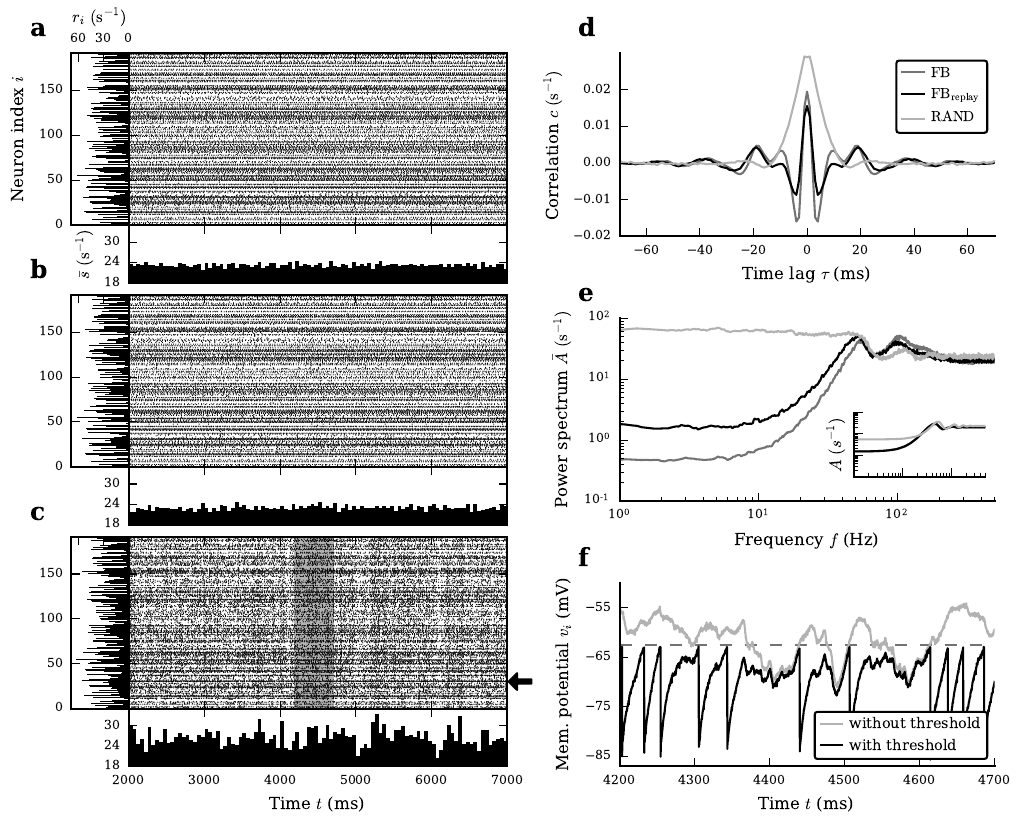}
\caption{
  Typical spiking and membrane-potential activity of a random inhibitory network of LIF neurons with intact and cut feedback loop emulated on the fully calibrated system.
  \textbf{(a--c)}
  Spiking activity (raster plots), population activity $\popactivity(t)$
  (horizontal histograms; bin size $\SI{50}{\milli\second}$)
  and time-averaged single-neuron firing rates $\rateindex$ (vertical histograms) in the 
  network with intact feedback (a) and for cases where the feedback loop is cut (b and c). 
  (a) Intact recurrent network (\fb{} scenario).
  (b) Population of mutually unconnected neurons receiving input spike trains identical to those in (a) (\fbrep{} scenario).
  (c) As in (b), but after randomization of presynaptic spike times (\rand{} scenario).
  \textbf{(d,e)}
  Population-averaged cross-correlation functions $\MakeLowercase{\crossspec}(\tau)$ (after offset subtraction) of pairs of spike trains (d) and power spectra $\poppowerspec(f)$ (e; log-log representation) of the population activity $\popactivity(t)$
  (cf.~horizontal histograms in (a--c)) for the \fb{} (dark gray), \fbrep{} (black) and \rand{} scenario (light gray).
  Inset in (e): Population-averaged power spectra $\powerspec(f)$ of individual single-cell spike trains (same scales as in main panel). 
  Correlation functions and spectra are averaged across $\Mtrials=100$ network realizations.
  \textbf{(f)}
  Membrane potential of a neuron in the \rand{} scenario (with firing rate of \SI{23.20}{\per\second} close to population average of \SI{23.24}{\per\second}; 
  see black arrow in (c)) with intact (black curve) and removed threshold (gray curve; \emph{free membrane potential}). 
  The threshold potential is marked by the horizontal dashed line. The time frame corresponds to the gray-shaded region in (c).
}
\label{fig:raster}
\end{figure*}

To quantify fluctuations in the population activity $\popactivity$ (\prettyref{fig:raster}a--c, horizontal histograms) we compute the power spectrum $\poppowerspec(f)$ of the population activity (\prettyref{fig:raster}e), which we scale with the duration $\runtime$ of the emulation.
Consequently, the population power spectrum $\poppowerspec(f)$, scaled by the population size, coincides with the time-averaged population activity $\meanrate$ for high frequencies: $\lim_{f \rightarrow \infty} \frac{1}{N}\poppowerspec(f) = \meanrate$ \citep{Tetzlaff08_2133}.

As a measure of pairwise correlations in the time domain (\prettyref{fig:raster}d), we compute the population-averaged cross-correlation function $\crossfunction(\tau)$ by Fourier transforming the population-averaged cross-spectrum $\crossspec(f)$ to time domain.

\section{Results}

In this study, we investigate the roles of shared input, feedback and heterogeneity on input and output correlations in random, sparse networks of inhibitory LIF neurons with conductance-based synapses (\prettyref{tab:nordlie}), implemented on the analog neuromorphic hardware chip \spikey{} (\prettyref{fig:hardware}).
Similarly to \cite{Tetzlaff12_e1002596}, we separate the contributions of shared input and feedback by studying
different network scenarios (\prettyref{fig:topo}): 
In the \fb{} case, we emulate the recurrent network with intact feedback loop (\prettyref{fig:topo}b) and record its spiking activity (\prettyref{fig:raster}a).  
In the \fbrep{} case (\prettyref{fig:topo}c), the feedback loop is cut and replaced by the activity recorded in the \fb{} network.  
Ideally, the input to each neuron in the \fbrep{} case should be identical to the input of the corresponding neuron in the \fb{} network.
As the replay of spikes and the resulting postsynaptic currents and membrane potentials are not perfectly reproducible on the \spikey{} chip, the neural responses in the \fb{} and in the \fbrep{} scenario are slightly different (compare Figures \ref{fig:raster}a and \ref{fig:raster}b).
In the \rand{} case (Figures \ref{fig:topo}d and \ref{fig:raster}c), we use the same setup as in the \fbrep{} case.  
However, the spike times in each presynaptic spike train are randomized.  
While the average presynaptic firing rates and the shared-input structure are exactly preserved in this scenario, the spatio-temporal correlations in the presynaptic spiking activity are destroyed.

Using this setup, we first demonstrate in \prettyref{sub:results_decorrelation} that active decorrelation by inhibitory feedback \cite{Renart10_587,Tetzlaff12_e1002596} is effective in heterogeneous networks with conductance-base synapses over a range of different network sizes.
In \prettyref{sub:results_heterogeneity}, we show that decreasing the level of heterogeneity by calibration of hardware neurons leads to an enhancement of this active decorrelation.

\subsection{Decorrelation by inhibitory feedback}
\label{sub:results_decorrelation}

The time-averaged population activities in the \fb{}, \fbrep{} and \rand{} scenarios are roughly identical (vertical histograms in Figures \ref{fig:raster}a--c; see also high-frequency power in \prettyref{fig:raster}e).  
In the \fb{} and \fbrep{} scenario, fluctuations in the population-averaged activity are small (horizontal histograms in \prettyref{fig:raster}a and b).  
The removal of spatial and temporal correlations in the presynaptic spike trains in the \rand{} case leads to a significant increase in the fluctuations of the population-averaged response activity (horizontal histogram in \prettyref{fig:raster}c).  
At low frequencies ($\le\SI{20}{\hertz}$), the population-rate power in the \fb{} and in the \rand{} case differs by about two orders of magnitudes (dark and light gray curves in \prettyref{fig:raster}e). 
This increase in low-frequency fluctuations in the \rand{} case is mainly caused by an increase in pairwise correlations in the spiking activity (\prettyref{fig:raster}d; the power spectra of individual spike trains [inset in \prettyref{fig:raster}e] are only marginally affected by a randomization of presynaptic spike times) \cite{Tetzlaff12_e1002596}.
In other words, shared-input correlations, i.e., those leading to large spike-train correlations in the \rand{} scenario, are efficiently suppressed by the feedback loop in the \fb{} case.

On the neuromorphic hardware, the replay of network activity is not perfectly reproducible (\prettyref{sub:methods_reprod}).
While the across-trial variability in membrane potentials is small, postsynaptic spikes are dithered by few milliseconds (\prettyref{fig:reprod}).
In the \fbrep{} case, the suppression of shared-input correlations by correlations in presynaptic spike trains is slightly less efficient as compared to the intact network (\fb{}).
The differences in the population-rate power spectra and in the spike-train correlations between the \fbrep{} and \rand{} case, respectively, are nevertheless substantial (solid black and light gray curves in \prettyref{fig:raster}d and e; note the logarithmic scale; for a detailed investigation of spike dither see \prettyref{app:sim} and \prettyref{appfig:sim_dither_power}).

\begin{figure*}
\includegraphics{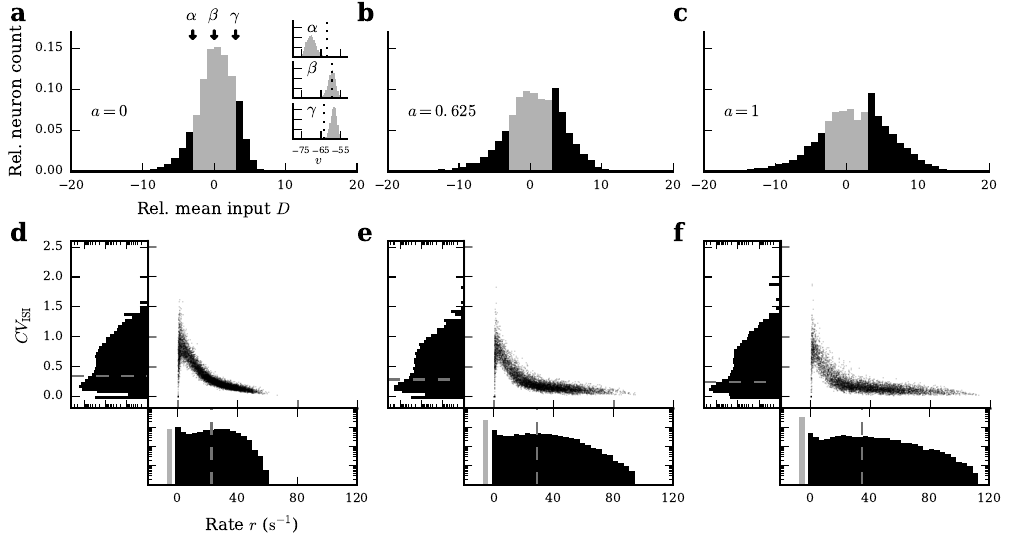}
\caption{
Modulation of network heterogeneity by leak-conductance calibration (see \prettyref{sub:methods_calib}).
Input (top row) and firing statistics (bottom row) in the intact recurrent networks (\fb{} scenarios) for
fully calibrated (a and d; $\calibfactor=0$), partially calibrated (b and e; $\calibfactor=0.625$) and uncalibrated neurons (c and f; $\calibfactor=1$).
\textbf{(a--c)}
Effect of calibration on input statistics.
Distributions of relative mean input $\dist=(\meanmem - \vthresh) / \stdmem$ 
(distance of time averaged free membrane potential $\meanmem$ from firing threshold $\vthresh$ in units of the standard deviation $\stdmem$)
across the population of neurons.
Gray areas in (a), (b) and (c) highlight $[-3,3]$ intervals, containing $74\%$, $53\%$ and $42\%$ of the total mass of the distribution, respectively.
Inset in (a): Distributions of free membrane potentials $\vm$ for three neurons $\alpha$, $\beta$ and $\gamma$ with $\dist = -3$, $\dist =0$ and $\dist =3$ (arrows in (a)), respectively.
Dotted lines mark threshold potentials that may vary due to fixed-pattern noise.
\textbf{(d--f)}
Effect of calibration on spike-train statistics.
Joint (scatter plots) and marginal distributions of single-neuron firing rates $\rate$ (horizontal histograms; log-linear scale) and coefficients of variation $\cvisi$ of inter-spike intervals (vertical histograms; log-linear scale).
Dashed lines mark mean of firing rate (\SI{22.6}{\per\second}, \SI{28.7}{\per\second}, \SI{34.8}{\per\second}) and $\cvisi$ distributions ($0.35$, $0.28$, $0.25$), respectively.
Gray bars (bottom panels) represent fractions of silent neurons.
Data obtained from $\Mtrials=50$ different network realizations.
Percentage of dead neurons: $8\%, 26\%, 37\%$.
}
\label{fig:rates_cv}
\end{figure*}

Note that the suppression of correlations and, hence, population-rate fluctuations by inhibitory feedback is restricted to low frequencies (here, to frequencies $<{}\SI{50}{\hertz}$; see \prettyref{fig:raster}e).
In the remainder of this study, we will quantify pairwise correlations by the low-frequency coherence in the range \SI{0.1}{}--\SI{20}{\hertz} (see \prettyref{sub:methods_decorr}). 
At higher frequencies, the population-rate power spectra in the \fb{}, \fbrep{} and \rand{} case are similar.
In \prettyref{fig:raster}e, the peaks at $\sim{}\SI{50}{\hertz}$ and higher harmonics result from the single-cell spike-train statistics (they are also visible in the single-cell spectra; see inset):
A large fraction of cells, in particular those firing at higher rates, generate regular spike trains with low inter-spike-interval (ISI) variability (cf.~\prettyref{fig:rates_cv}d--f).
These (fast spiking) cells contribute maxima to the spike-train spectra at frequencies close to their firing rates (and higher harmonics).
The structure of the population-rate spectra at higher frequencies ($\ge{}\SI{50}{\hertz}$) is reproduced using surrogate data where the ISI distributions of the individual neurons (and, hence, their firing rates and ISI variability) are preserved,
but serial ISI correlations and cross-correlations between spike trains are destroyed (data not shown).

\begin{figure*}
\includegraphics{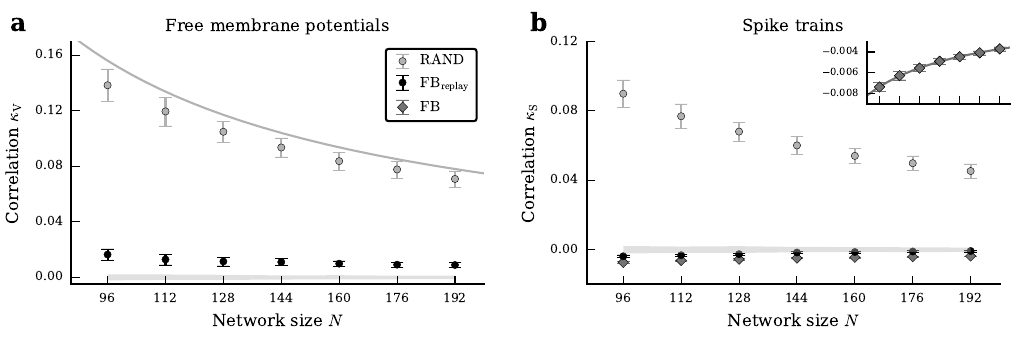}
\caption{
  Dependence of population-averaged input correlations \textbf{(a)} and spike-train correlations \textbf{(b)} on the network size $\netsize$, for
  the intact network (\fb{}, dark gray diamonds), the \fbrep{} (black circles) and the \rand{} (light gray circles) case (fixed in-degree $\noinputs=15$).
  Symbols and error bars denote mean and one standard deviation, respectively, across $\Mtrials=100$ network realizations (error bars are partly covered by markers).
  The gray curve in (a) depicts shared-input correlations in a homogeneous network (\prettyref{eq:shared_input}).
  The inset in (b) shows a magnified view of the spike-train correlations in the \fb{} case (dark gray diamonds) with a power-law fit $\sim \netsize^{-1}$ (dark gray curve).
  The light gray horizontal band represents mean $\pm$ three standard deviations of (spurious) correlations in surrogate data where correlations were removed.
  Note that free membrane potentials cannot be recorded in the \fb{} case (see \prettyref{sec:methods}).
  Hence, there are no gray diamonds in (a).
}
\label{fig:corr_n}
\end{figure*}

In the \rand{} case, presynaptic spike-train correlations were removed,
and hence input (i.e., free-membrane-potential) correlations are exclusively determined by the number of shared presynaptic sources (\prettyref{eq:shared_input}).
If the in-degree $\noinputs$ is fixed, input correlations will decrease with network size $\netsize$ (\prettyref{eq:shared_input}, light gray curve and symbols in \prettyref{fig:corr_n}a). 
In purely inhibitory networks with intact feedback loop (\fb{} scenario), correlations in presynaptic spike trains are on average significantly smaller than zero (dark gray diamonds in \prettyref{fig:corr_n}b, \citep{Tetzlaff12_e1002596}),
and largely cancel the positive contribution from shared-input correlations.
Average input correlations are therefore significantly reduced (black symbols in \prettyref{fig:corr_n}a).
As both shared-input and spike-train correlations scale with the inverse of the network size ($\netsize^{-1}$; light gray curve in \prettyref{fig:corr_n}a and inset in \prettyref{fig:corr_n}b, respectively) \citep{DeLaRocha07_802},
this suppression of correlations in the \fb{} (and \fbrep{}) case is observed for all investigated network sizes $N$.
Note that output correlations are negative even though input correlations are positive.
This effect is predicted by theory and also observed in linear network models as well as LIF-network simulations on conventional computers (see \prettyref{fig:sim}, \prettyref{app:sim}, \ref{app:linear} and \prettyref{sec:discussion}).

\subsection{Effect of heterogeneity on decorrelation}
\label{sub:results_heterogeneity}

In neural networks implemented in analog neuromorphic hardware, neuron and synapse parameters vary significantly across the population of cells (fixed-pattern noise; see \prettyref{sub:methods_hardware}).
For a population of mutually unconnected neurons with distributed parameters, injection of a constant (supra-threshold) input current leads to a distribution of response firing rates (\prettyref{fig:calib}).
In this study, we consider the width of this firing-rate distribution as a representation of neuron heterogeneity.
It is systematically varied by calibration of leak conductances.
The extent of heterogeneity is quantified by the calibration parameter $a$ ($a=1$ and $a=0$ correspond to the uncalibrated and the fully calibrated system, respectively; for details, see \prettyref{sub:methods_calib}).
For an unconnected population of neurons subject to constant input, the width of the firing-rate distribution increases monotonically with $a$.

As shown in \prettyref{fig:rates_cv}, the level of heterogeneity (i.e., the calibration state $a$) is clearly reflected in the activity of the recurrent network (\fb{} case).
Both the width of the distribution of mean free membrane potentials (\prettyref{fig:rates_cv}a--c) as well as the width of the firing-rate distribution increases with $a$ (\prettyref{fig:rates_cv}d--f; horizontal histograms). %% ``Both A and B'' or ``both A as well as B'' without comma.
In the uncalibrated system ($a=1$), a substantial fraction of neurons is predominantly driven by constant supra-threshold input currents and therefore generates highly regular spike trains ($\cvisi\approx{}0$) with high firing rates ($\rate > \SI{60}{\second}^{-1}$).
Simultaneously, about $37\%$ of the neurons are silent ($\rate = \SI{0}{\second}^{-1}$).
Neurons with intermediate firing rates ($\SI{0}{\second}^{-1} < \rate < \SI{15}{\second}^{-1}$), however, show quite irregular activity ($\cvisi > 0.5$).
After calibration, the firing-rate distribution is narrowed.
For $a=0$, the fraction of silent neurons is reduced to about $8\%$.
Maximum rates are limited to $<\SI{66}{\second}^{-1}$.
Note that our calibration routine compensates only for the distribution of neuron parameters, but not for the heterogeneity in synapse properties (synaptic weights, synaptic time constants; see \prettyref{sec:discussion}).
Even for the fully calibrated network ($a=0$), the firing-rate distribution is therefore still broad.
In the \rand{} case, we obtain similar firing-rate and ISI statistics as in the \fb{} case (\prettyref{app:statistics_rand}).

\begin{figure*}
\includegraphics{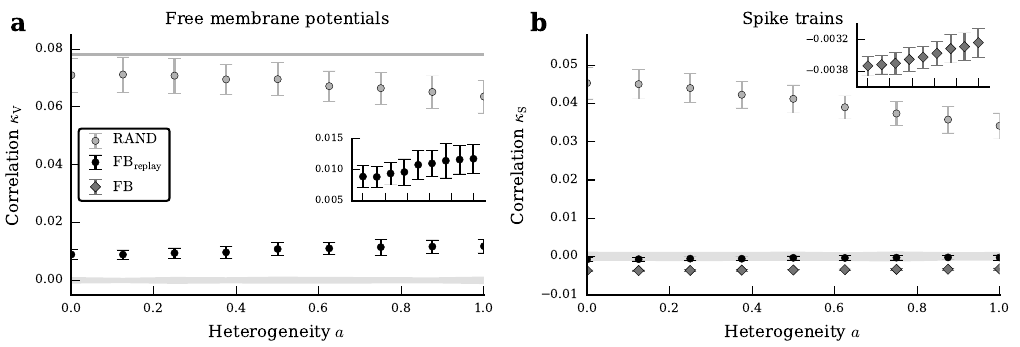}
\caption{
  Dependence of population-averaged input correlations \textbf{(a)} and spike-train correlations \textbf{(b)} on the heterogeneity of the neuromorphic substrate
  for the intact network (\fb{}, dark gray diamonds), the \fbrep{} (black circles) and the \rand{} (light gray circles) case.
  Symbols and error bars denote mean and one standard deviation, respectively, across $\Mtrials=100$ network realizations (error bars are partly covered by markers).
  The gray line in (a) depicts shared-input correlations in a homogeneous network (\prettyref{eq:shared_input}).
  Insets in (a) and (b) depict magnified views of input correlations in the \fbrep{} case and spike-train correlations in the \fb{} case, respectively.
  The light gray horizontal band represents mean $\pm$ three standard deviations of (spurious) correlations in surrogate data where correlations were removed.
  Note that free membrane potentials cannot be recorded in the \fb{} case (see \prettyref{sec:methods}).
  Hence, there are no gray diamonds in (a).
}
\label{fig:corr_calib}
\end{figure*}

For all levels of heterogeneity attainable by our calibration procedure ($a\in[0,1]$), input and output correlations are significantly suppressed by the recurrent-network dynamics (cf.\ black and dark gray vs.\ light gray symbols in \prettyref{fig:corr_calib}).
In a homogeneous, random (Erd\H{o}s-R\'{e}nyi) network with fixed in-degree $K$ and linear sub-threshold dynamics, the contribution of shared input to the input (free-membrane-potential) correlation is given by the network connectivity $\noinputs/\netsize$ \citep[\prettyref{eq:shared_input}; reference][]{Tetzlaff12_e1002596} (thin light gray curves in \prettyref{fig:corr_n}a and \prettyref{fig:corr_calib}a).
Nonlinearities in synaptic and/or spike-generation dynamics \citep{DeLaRocha07_802} as well as heterogeneity in neuron (and synapse) parameters lead to a suppression of this contribution (\prettyref{eq:shared_input_hetero}, \citep{Padmanabhan10_1276}).
Here, we refer to this type of decorrelation as \emph{feedforward decorrelation}.
In fact, in our setup the input and spike-train correlations in the \rand{} case decrease with increasing heterogeneity (light gray symbols in \prettyref{fig:corr_calib}).
Even in the fully calibrated case input correlations are slightly smaller than $\noinputs/\netsize$ (gray symbols vs.\ thin gray curve in \prettyref{fig:corr_calib}a) due to remaining heterogeneities.
Spike-train correlations decrease slightly faster with increasing heterogeneity than input correlations.
These observations indicate that both the synaptic integration and spike generation are affected by heterogeneities on hardware.
To illustrate the different effects of heterogeneity in synaptic integration and spike generation, we performed network simulations on conventional computers where we distributed either firing thresholds (\prettyref{fig:sim}) or synaptic weights (\prettyref{appfig:sim_weights}).
In the \rand{} case, the decrease of input correlations on the hardware can be attributed to an increase in heterogeneity in parameters affecting synaptic integration (compare light gray symbols in \prettyref{fig:corr_calib}a and \prettyref{appfig:sim_weights}a).
Heterogeneity in spike thresholds, in contrast, does not affect input correlations in the \rand{} scenario, but strongly reduces spike-train correlations (light gray symbols in \prettyref{fig:sim}).
Overall, feedforward decorrelation, i.e.~the suppression of correlations in the \rand{} case, becomes more effective in networks with heterogeneous cell parameters.

Despite this enhancement of feedforward decorrelation, input and output correlations increase with the level of heterogeneity in the presence of an intact feedback loop (\fb{} and \fbrep{} scenarios; black and dark gray symbols in \prettyref{fig:corr_calib}).
We attribute this effect to a weakening of the effective feedback loop in the recurrent circuit:
In heterogeneous networks with broad firing-rate distributions, neurons firing with low or high rates, corresponding to mean inputs far below or far above firing threshold (see \prettyref{fig:rates_cv}a--c), are less sensitive to input fluctuations than moderately active neurons (see \prettyref{appfig:weff_ratehist}).
Hence, they contribute less to the overall feedback.
In consequence, feedback decorrelation is impaired by heterogeneity (see also \prettyref{sec:discussion}).

\section{Discussion}

\label{sec:discussion}

We have shown that inhibitory feedback effectively suppresses correlations in heterogeneous recurrent neural networks of leaky integrate-and-fire (LIF) neurons with nonlinear subthreshold dynamics, emulated on analog neuromorphic hardware (\spikey{}; \citep{Schemmel06_1,Pfeil13_11}).
Both input and output correlations are substantially smaller in networks with intact feedback loop (\fb{}), as compared to the case where the feedback is replaced by randomized input while preserving the connectivity structure and presynaptic firing rates (\rand{}).
Our results hence show that active decorrelation of network activity by inhibitory feedback \citep{Renart10_587,Tetzlaff12_e1002596} is a general phenomenon which can be observed in realistic, highly heterogeneous networks with nonlinear interaction and sufficiently strong negative feedback.
Moreover, the study serves as a proof-of-principle that network activity can be efficiently decorrelated even on heterogeneous hardware, which can be exploited in functional applications, e.g., in the neuromorphic algorithms developed by \citet{Pfeil13_11} and \citet{Schmuker14_2081}.

Functional neural architectures often rely on stochastic dynamics of its constituents or on some form of background noise \citep[see, e.g.,][]{Maass14_860,Pfeil13_11, Schmuker14_2081}.
Deterministic recurrent neural networks with inhibitory feedback could provide decorrelated noise to such functional networks, both in artificial as well as in biological substrates.
In neuromorphic hardware applications, these ``noise networks'' could thereby replace conventional random-number generators and avoid a costly transmission of background noise from a host computer to the hardware substrate (which may be particularly relevant for mobile applications with low power consumption; see \prettyref{app:power}).
It needs to be investigated, however, how well functional stochastic circuits perform in the presence of such network-generated noise.

Partial calibration of hardware neurons allowed us to modulate the level of network heterogeneity and, therefore, to systematically study its effect on correlations in the network activity.
The analysis revealed two counteracting contributions:
As shown in previous studies \citep[e.g.,][]{Padmanabhan10_1276}, neuron heterogeneity decorrelates (shared) feedforward input (feedforward decorrelation).
On the other hand, however, heterogeneity impairs feedback decorrelation (see next paragraph).
In our network model, this weakening of feedback decorrelation is the dominating factor.
Overall, we observed a slight increase in correlations with increasing level of heterogeneity.
We cannot exclude that feedforward decorrelation may play a more significant role for different network configurations (e.g., different connection strengths or network topologies, different structure of external inputs, different types of heterogeneity).
Our study demonstrates, however, that heterogeneity is not necessarily suppressing correlations in recurrent systems.
In this context, it would be interesting to investigate the interplay of signal and noise correlations in the presence of network heterogeneities in recurrent systems.
We leave this intriguing topic to future studies.

As shown in \citep{Tetzlaff12_e1002596}, feedback decorrelation in recurrent networks becomes more (less) efficient with increasing (decreasing) strength of the \emph{effective} negative feedback.
For networks of spiking neurons, the effective connection strength $w_{ij}$ between two neurons $j$ and $i$ corresponds to the total number of extra spikes emitted by neuron $i$ in response to an additional input spike generated by neuron $j$ (see, e.g., \citep{London10_123}).
Assuming that the effect of a single additional input spike is small, the effective connectivity can be obtained by linear-response theory \citep{DeLaRocha07_802}.
Note that the effective weights $w_{ij}$ depend on the working point, i.e., the average firing rates of all pre- and postsynaptic neurons (mathematically, $w_{ij}$ is given by the derivative of the stationary response firing rate $r_i=\sub{\phi}{i}(r_1,\ldots,r_j,\ldots,r_N)$ of neuron $i$ with respect to the input firing rate $r_j$, evaluated at the working point; for details, see \citep{Tetzlaff12_e1002596}).
Neurons firing at very low or very high rates are typically less sensitive to input fluctuations than neurons firing at intermediate rates (due to the shape of the response function $\sub{\phi}{i}(r_1,\ldots,r_N)$).
Their dynamical range is reduced.
In consequence, they hardly mediate feedback in a recurrent network.
In heterogeneous networks with broad distributions of firing rates, the number of these insensitive neurons is increased.
Hence, the effective feedback is weakened (see \prettyref{app:weff}).
We can qualitatively reproduce this effect of heterogeneity on correlations in recurrent networks (\fb{} case) by means of a simplified linear rate model where increasing heterogeneity is described as a decrease in the effective-weight amplitudes (see \prettyref{app:linear}).
A more quantitative analysis requires an explicit mapping of the synaptic weights in the LIF-neuron network to the effective weights of the linear model (as in, e.g., \citep{Grytskyy13_131}) in the presence of distributed firing rates.
We commit this task to future studies.
Note that the rate dependence of the effective weights and the resulting effects on correlations are consistent with our observation that neuron pairs with very low firing rates exhibit spike-train correlations close to zero, whereas pairs with high firing rates are positively correlated (see \prettyref{app:corrmatrix}).
Pairs with one neuron firing at an intermediate rate often exhibit negative spike-train correlations.
As shown in \citep{Renart10_587,Tetzlaff12_e1002596}, these negative spike-train correlations are essential for compensating the positive contribution of shared inputs to the total input correlation (at least in purely inhibitory networks).
Narrowing the firing rate distribution (e.g., by calibration of hardware neurons) increases the number of neurons contributing to the negative feedback, which,
in turn, leads to more neuron pairs with negative spike-train correlations and, therefore, to smaller overall correlations.

Seemingly contrary to our findings, \citet{Bernacchia13_1732} report a decrease in correlations with increasing level of heterogeneity.
The results of their study are obtained for a linear network model, which can be considered the outcome of the linearization procedure described above.
Hence, the connectivity of their model corresponds to an effective connectivity (see above).
Their study neglects the rate (working-point) dependence of the effective weights and can therefore not account for the effect of firing-rate heterogeneity.
In \citep{Bernacchia13_1732}, heterogeneity is quantified by the variance of the (effective) weight matrix (Equations 2.2 and 2.4 in \citep{Bernacchia13_1732}).
For sparse connectivity matrices (with a large number of zero elements), the variance of the weight matrix reflects not only the width of the non-zero-weight distribution, but also its mean (Equation 2.4 in \citep{Bernacchia13_1732}).
For networks of nonlinear spiking neurons, heterogeneities in neuron and/or synapse parameters broadens the distribution of non-zero effective weights, but may simultaneously reduce its mean (see above, \prettyref{app:weff}, and \citep{Roxin11_16217,Grytskyy13_131}).
Hence, the variance of the full weight matrix may decrease (for illustration, see \prettyref{appfig:effective_weight_dist}).
In other words, increasing heterogeneity in the nonlinear system may correspond to decreasing heterogeneity in the linearized system.
A direct test of this hypothesis requires an explicit linearization of the nonlinear heterogeneous system.

\begin{figure*}
\centering
\includegraphics{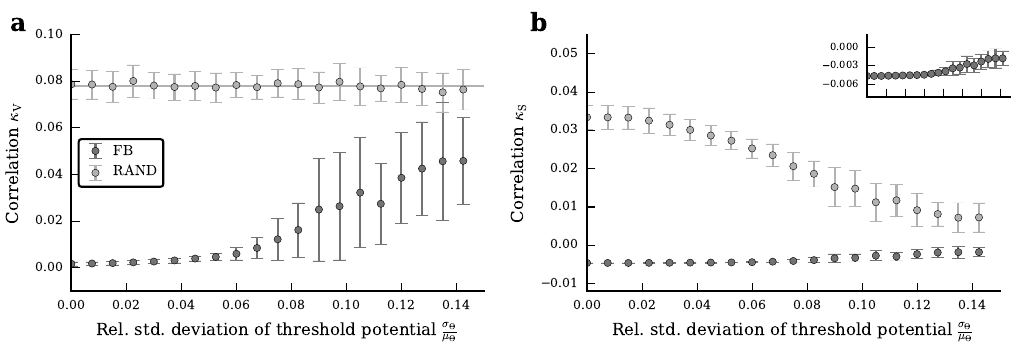} % N=192

\caption{
  Dependence of population-averaged input correlations \textbf{(a)}, and spike-train correlations \textbf{(b)} on the width of threshold distributions in networks simulated with NEST \citep{Gewaltig_07_11204} and PyNN \citep{Davison08_11}, for the intact network (\fb{}, dark gray circles) and the \rand{} (light gray circles) case.
  Symbols and error bars denote mean and standard deviation, respectively, across $\Mtrials=30$ network realizations (error bars are partly covered by markers).
  The gray line in (a) depicts shared-input correlations in a homogeneous network (\prettyref{eq:shared_input}).
  The inset in (b) shows a magnified view of the spike-train correlations in the \fb{} case.
  Note that in simulations the \fbrep{} is identical to the \fb{} case, and is hence not shown.
  For details see \prettyref{app:sim}.
}
\label{fig:sim}
\end{figure*}

The results of this study were obtained by network emulations on analog neuromorphic hardware.
We reproduced the main findings by means of conventional computer simulations of LIF-neuron networks with distributed firing thresholds (see \prettyref{fig:sim}).
The focus on threshold heterogeneity allows us to isolate the effect of firing-rate distributions on correlations.
It does not affect shared-input correlations (see \prettyref{sub:methods_decorr}).
Although networks simulated on conventional computers and those emulated on the neuromorphic hardware differ in several respects (e.g., in the exact implementation of heterogeneity or the synapse model; compare \prettyref{apptab:sim_nordlie} and \ref{apptab:sim_nordlie_params} to \prettyref{tab:nordlie} and \ref{tab:nordlie_params}, respectively), the qualitative results are very similar:
In networks with intact feedback loop, input and output correlations are substantially reduced (as compared to the case where the feedback is replaced by randomized input), but increase with the extent of heterogeneity. 
As predicted by the theory for homogeneous inhibitory networks, we observe positive input correlations and negative output correlations (see Equation $21$ in \cite{Tetzlaff12_e1002596} and in the paragraph which follows; see also \cite{Helias13_023002} and \prettyref{app:linear}).
Further, note that heterogeneity in neuron parameters does not ``average out'' in larger networks. 
Upscaling the network size by a factor of $25$ ($\netsize=4800$, in-degree $K=375$) yields smaller spike-train correlations, but the qualitative results are similar to those obtained for the smaller network ($N=192$, $K=15$) emulated on the \spikey{} chip (compare \prettyref{fig:corr_calib} to \prettyref{appfig:sim}).

In networks with intact feedback loop (\fb{} and \fbrep{} scenarios), the precise spatio-temporal structure of spike trains arranges such that the self-consistent input and output correlations are suppressed.
Perturbations of this structure in the local input typically lead to an increase in correlations \citep{Tetzlaff12_e1002596}.
In this study, we demonstrate this by replaying spiking activity after randomization of spike times, i.e., by replacing the time of each input spike by a random number uniformly drawn from the full emulation time interval $[0,T)$ (\rand{} case).
However, even subtle modifications of input spike trains, such as random dither of spike times by few milliseconds, lead to an increase of correlations.
On the neuromorphic hardware, replay of spike trains is not entirely reproducible (see \prettyref{sub:methods_reprod} and \prettyref{app:control}).
Hence, spike-train correlations measured in the \fbrep{} mode are slightly larger than in the \fb{} case.
We would expect the same effect on the input side (free membrane potentials).
Due to hardware limitations, however, we can measure input correlations only in replay mode (\fbrep{} or \rand{}), but not in the fully connected network (\fb{}).
Therefore, all reported input correlations are likely to be slightly overestimated.
In conventional network simulations, we mimicked the effect of unreliable replay by input-spike dithering and, indeed, find a gradual increase in input and output correlations (see \prettyref{appfig:sim_dither}).
These results seem to be contrary to the study by \citet{Rosenbaum11_1261}, in which synaptic noise leads to a decrease of output correlations in a feedforward scenario.
In our case, spike-train correlations, which suppress shared-input correlations, are removed by dithering spikes, thereby increasing correlations on the output side.
In \citep{Rosenbaum11_1261}, in contrast, spike-train correlations are always zero, and shared-input correlations are decreased by synaptic failure, explaining the decreased output correlations.
We attribute this contradiction to the missing feedback loop in their system, and expect correlations to increase in recurrent networks subject to similar perturbations.

Despite the imperfect replay of input spikes, the decorrelation effect is clearly visible in hardware emulations, both on the input and on the output side.
The reproducibility of emulations on neuromorphic hardware could be improved by stabilizing the environment of the system, e.g., the chip temperature or the support electronics (under development).
Analog hardware, however, will never reach the level of reproducibility of digital computers.
But note that, similar to analog hardware, biological neurons exhibit a considerable amount of trial-to-trial variability, even under controlled in-vitro conditions \citep{Mainen95_1503}.
So far, the details of how neuronal noise, for example, stochastic synapses (spontaneous postsynaptic events, stochastic spike transmission, synaptic failure \citep{Ribrault11_375}), affects correlations in recurrent neural circuits remain unclear.

Although different \spikey{} chips exhibit different realizations of fixed-pattern noise, they show a comparable extent of heterogeneity and yield results which are qualitatively similar to those presented in this article (\prettyref{app:chips}).

We have shown that negative feedback in recurrent circuits can efficiently suppress correlations, even in highly heterogeneous systems such as the analog neuromorphic architecture \spikey{}.
Correlations can be further reduced by minimizing the level of network heterogeneity.
In this study, we reduced the level of heterogeneity through calibration of neuron parameters in the unconnected case (see \prettyref{sub:methods_calib}).
The calibration could, in principle, be improved by calibrating neuron (and possibly synapse) parameters in the full recurrent network.
Such calibration procedures are however time consuming and cumbersome.
In biological substrates, homeostasis mechanisms \citep{Marder06_563,oleary14_809} keep neurons in a responsive regime and reduce the level of firing-rate heterogeneity in a self-regulating manner.
Future neuromorphic devices could mimic this behavior, thereby reducing the necessity of time consuming calibration procedures.
Alternatively, the analog circuits could be optimized to reduce fixed-pattern noise.
This would likely require the allocation of more chip resources, hence reducing the network size per chip area.

For simplicity, this work focuses on purely inhibitory networks (as in \citep{Mar99_10450}).
This demonstrates that decorrelation by inhibitory feedback does not rely on a dynamical balance between excitation and inhibition (note that the external ``excitatory'' drive is constant in our model) \citep{Tetzlaff12_e1002596,Helias14}.
Previous studies have shown that, for the homogeneous case, decorrelation by inhibitory feedback is a general phenomenon, which also occurs in excitatory-inhibitory networks, provided the overall inhibition is sufficiently strong (which is typically the case to ensure stability) \citep{Renart10_587,Tetzlaff12_e1002596,Bernacchia13_1732,Helias14}.
For the heterogeneous case, computer simulations of excitatory-inhibitory networks show qualitatively the same results as purely inhibitory networks (compare \prettyref{fig:corr_calib} to \prettyref{appfig:sim_ei}),
confirming that our results generalize to the case of mixed excitatory-inhibitory coupling.

Similar to our study, \citet{Giulioni12_149} use a theory-guided approach to implement, verify and investigate network dynamics on analog neuromorphic hardware.
In their study, an attractor network is implemented that is inspired by a mean-field model.
Due to heterogeneities in synaptic efficacies on the hardware, stability analysis of attractor states requires the authors to measure \emph{effective response functions} of populations of hardware neurons.
To this end, they replace recurrent connections in one population of neurons by external input.
This allows them to measure the firing rate of the population as a function of the external input,
while the activity of the population is in equilibrium with that of other recurrently connected populations.
This study represents another example illustrating that investigations of actual hardware emulations are a prerequisite for successful application of analog neuromorphic hardware.

\begin{figure}
\includegraphics{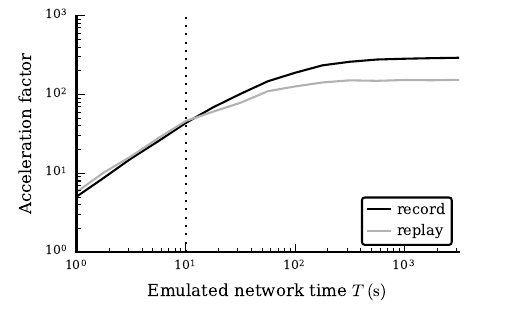}
\caption{
Acceleration factor as a function of emulated network time $\runtime$ for the record (black) and the replay case (gray).
The acceleration factor is defined as the ratio between the emulated network time $\runtime$ (in biological time) and the execution time (wall clock time). 
In the record case, a network realization is generated on the host computer and uploaded to the chip.
During the subsequent emulation, spike trains are recorded.
In the replay case, spikes are replayed and the membrane potential of one neuron is recorded with full sampling frequency (\SI{9.6}{\kilo\hertz}).
The execution time covers the full data flow from a network description in \pynn{} to the emulation on the \spikey{} system and back to the network representation in \pynn{}.
The time-averaged population firing rate is $\meanrate = \SI{23.7 \pm 0.2}{\per\second}$.
The vertical dashed line depicts the runtime used in this study.
The hardware system has to be initialized once before usage ($<\SI{1}{\second}$), which is not considered here.
}
\label{fig:profiling}
\end{figure}

This study demonstrates that the \spikey{} system has matured to a level that permits its use as a tool for neuroscientific research.
For the results presented in this study, we recorded in total $10^{11}$ membrane-potential and spike-train samples, representing more than $100$ days of biological time.
Due to the $10^4$-fold acceleration of the \spikey{} chip, this corresponds to less than $15$ minutes in the hardware-time domain.
              %membrane: 1600h = 5.5 * 10^10 Samples @ 9.6kHz
              %spikes: 12,5h = 2.6 * 10^8 Spikes @ 30Hz (<-not true, because smaller networks)
              %membrane: 160h
              %membrane + spikes: 160h
              %spikes: 0,4h
              %spikes: 0,6h
              %spikes: 1,25h
              %membrane: 667h
              %spikes: 3,5h
              %membrane: 21,3h
              %spikes: 0,2h 
Interfacing the hardware system, however, reduces the acceleration to an approximately $50$-fold speed-up (\prettyref{fig:profiling}).
The translation between the network description and its hardware representation claims the majority of execution time, more than the network emulation and the transfer of data to and from the hardware system together.
Encoding and decoding spike times on the host computer is particularly expensive.
Obviously, the system could be optimized by processing the data directly on the hardware, or by choosing a data representation which is closer to the format used on the \spikey{} chip, but this would impair user-friendliness, and hence, the effectiveness of prototyping.
While the \spikey{} system permits the monitoring of the spiking activity of all neurons simultaneously, access to the membrane potentials 
is limited to a single (albeit arbitrary) neuron in each emulation run.
Monitoring of membrane potentials of a population of $n$ neurons therefore requires $n$ repetitions of the same emulation.
Extending the hardware system to enable access to the membrane potentials of at least two neurons simultaneously would allow for a direct observation of input correlations in the intact network (and thereby avoid problems with replay reproducibility; see above) and reduce execution time (the \spikey{} chip itself permits recording of up to eight neurons in parallel, the support electronics, however, does not). %% note: ``electronics'' is used with a singular verb
While the \spikey{} system does not significantly outperform conventional computers in terms of computational power, emulations on this system are much more energy efficient (\prettyref{app:power}).
A substantial increase of computational power is expected for large systems exploiting the scalability of this technology without slow-down \citep{Schemmel10_1947}.

\vspace{1em} %strange behavior here! beforeskip from revtex section: 0.8cm plus1ex minus.2ex

\begin{acknowledgments}
We would like to thank Matthias Hock, Andreas Hartel, Eric M{\"u}ller and Christoph Koke for their support in developing and building the \spikey{} system, as well as Mihai Petrovici and Sebastian Schmitt for fruitful discussions.
This research is partially supported by the Helmholtz Association portfolio theme SMHB, the J{\"u}lich Aachen Research Alliance (JARA), EU Grant 269921 (BrainScaleS), and EU Grant 604102 (Human Brain Project, HBP).
We acknowledge financial support by Deutsche Forschungsgemeinschaft and Ruprecht-Karls-Universit{\"a}t Heidelberg within the funding programme Open Access Publishing.
All network simulations carried out with NEST (http://www.nest-simulator.org).

\end{acknowledgments}

\appendix
\section{Network description}
\begingroup
\squeezetable
\begin{table*}

\setlength{\columnwidthleft}{0.2\fullfigwidth}
\setlength{\columnwidthmiddle}{0.2\fullfigwidth}

\begin{tabularx}{\fullfigwidth}{|p{\columnwidthleft}|X|}
  \hline\modelhdr{2}{A}{Model summary}\\\hline
  Populations    & One (inhibitory) \\
  \hline
  Topology       & None \\
  \hline
  Connectivity   & Random convergent connections (fixed in-degree) \\
  \hline
  Neuron model   & Leaky integrate-and-fire (LIF), fixed firing threshold, fixed absolute refractory time \\
  \hline
  Channel models & None \\
  \hline
  Synapse model  & Exponentially decaying conductances, fixed delays \\
  \hline
  Plasticity     & None \\
  \hline
  External input & Resting potential higher than threshold (= constant current) ($\vrest>\vthresh$) \\
  \hline
  Measurements   & Spikes and membrane potentials from all neurons \\
  \hline
  Other          & No autapses, no multapses \\
  \hline
\end{tabularx} \\

\begin{tabularx}{\fullfigwidth}{|p{\columnwidthleft}|p{\columnwidthmiddle}|X|}
  \hline\modelhdr{3}{B}{Populations}\\\hline
  \bf Name & \bf Elements & \bf Size \\
  \hline
  I & LIF neuron & $\netsize$ \\
  \hline
\end{tabularx} \\

\setlength{\columnwidthleft}{0.08\fullfigwidth}
\setlength{\columnwidthmiddle}{0.08\fullfigwidth}

\begin{tabularx}{\fullfigwidth}{|p{\columnwidthleft}|p{\columnwidthmiddle}|X|}
  \hline\modelhdr{3}{C}{Connectivity}\\\hline
  \bf Source & \bf Target & \bf Pattern \\
  \hline
  I & I & Random convergent connections, in-degree $\noinputs$ \\
  \hline
\end{tabularx} \\

\setlength{\columnwidthleft}{0.2\fullfigwidth}
\setlength{\columnwidthmiddle}{0.2\fullfigwidth}

\begin{tabularx}{\fullfigwidth}{|p{\columnwidthleft}|X|}
  \hline\modelhdr{2}{D}{Neuron and synapse model}\\\hline
  Type & Leaky integrate-and-fire, exponentially decaying conductances \\
  \hline
  Subthreshold dynamics &
    Subthreshold dynamics ($t \not\in (\ti, \ti + \tauref)$): \newline
    \hspace*{1em} $\cmem \frac{\text{d}}{\text{d}t} \vm(t) = -\gleak (\vm(t) - \vrest) - \ginh(t) (\vm(t) - \vinh)$ \newline
    Reset and refractoriness ($t \in (\ti, \ti + \tauref)$): \newline
    \hspace*{1em} $\vm(t) = \vreset$ \newline
    This model is emulated by analog circuitry on the \spikey{} chip \citep{Indiveri11_00073}. \\
  \hline
  Conductance dynamics &
    For each presynaptic spike at time $\ti$ ($t > \ti + d$): \newline
    \hspace*{1em} $\ginh(t) \approx \weight \exp(-\frac{t - \ti - \delay}{\tausyn})\Theta(t)$, with $\weight = \weighthw \gmax$ and Heaviside function $\Theta(t)$. \newline
    This model is emulated by analog circuitry on the \spikey{} chip \citep{Pfeil13_11}. \\
  \hline
  Spiking &
    If $\vm(\ti-) < \vthresh \wedge \vm(\ti+) \ge \vthresh$: \newline
    \hspace*{1em} emit spike with time stamp $\ti$ \\
  \hline
\end{tabularx}

\caption{
  Description of the network model (according to \citep{Nordlie-2009_e1000456}).
  \label{tab:nordlie}
}

\end{table*}
\endgroup

\begingroup
\squeezetable
\begin{table*}

\setlength{\columnwidthleft}{0.12\fullfigwidth}
\setlength{\columnwidthmiddle}{0.35\fullfigwidth}

\begin{tabularx}{\textwidth}{|p{\columnwidthleft}|p{\columnwidthmiddle}|X|}
  \hline\parameterhdr{3}{B}{Populations}\\\hline
  \bf Name & \bf Values & \bf Description\\\hline
  $\netsize$  & $\{96, 112, 128, 144, 160, 176, \textbf{192}\}$ & network size \\
  \hline\parameterhdr{3}{C}{Connectivity}\\\hline
  \bf Name & \bf Values & \bf Description\\\hline
  $\noinputs$ & $15$                          & number of presynaptic partners \\

  \hline\parameterhdr{3}{D}{Neuron}\\\hline
  \bf Name & \bf Values & \bf Description\\\hline
  $\vreset$     & \textcolor{graytarget}{\SI{-80}{\milli\volt}}          & reset potential \\
  $\vrest$      & \textcolor{graytarget}{\SI{-52}{\milli\volt}}          & resting potential \\
  $\vthresh$    & \textcolor{graytarget}{\SI{-62}{\milli\volt}}          & firing threshold \\
  $\vinh$       & \textcolor{graytarget}{\SI{-80}{\milli\volt}}          & inhibitory reversal potential \\
  $\tauref$     & \textcolor{graytarget}{\SI{1}{\milli\second}}          & refractory period \\
  $\cmem$       & \textcolor{graytarget}{\SI{0.2}{\nano\farad}}          & membrane capacitance \\
  $\gleak$      & {\it tuned during calibration process}                 & leak conductance \\
  $\taumemeff$  & uncalibrated: $10.50(5.94, 17.90)$\SI{}{\milli\second} & effective membrane time constant \\
                & calibrated: $10.77(6.22, 18.30)$\SI{}{\milli\second}   & \\

  \hline\parameterhdr{3}{D}{Synapse}\\\hline
  \bf Name     & \bf Values                                            & \bf Description\\\hline
  $\gmax$       & in the order of $\SI{1}{\nano\siemens}$               & conductance amplitude \\
  $\weighthw$   & $3$                                                   & synaptic weight (in hardware values $\in [0, 15]$) \\
  $\vmaxeff$    & uncalibrated: -$5.57(-12.50,-2.62)$\SI{}{\milli\volt} & effective post-synaptic potential prefactor \\
                & calibrated: -$6.05(-14.37,-2.92)$\SI{}{\milli\volt}   & \\
  $\tausyneff$  & uncalibrated: $3.78(2.52, 5.58)$\SI{}{\milli\second}  & effective synaptic-current time constant \\
                & calibrated: $4.17(2.76, 6.18)$\SI{}{\milli\second}    & \\
  $\delayeff$   & uncalibrated: $2.16(1.91, 2.48)$\SI{}{\milli\second}  & effective synaptic delay \\
                & calibrated: $2.26(1.93, 2.55)$                        & \\
  %

  %\hline\parameterhdr{3}{D}{Input}\\\hline
  %\bf Name & \bf Values & \bf Description\\\hline
  %& & $\vrest > \vthresh$ \\
  %%
  %\hline

  %\hline\parameterhdr{3}{Other}{Emulation}\\\hline
  %\bf Name & \bf Values & \bf Description\\\hline
  %%
  %$\runtime$  & \SI{5}{\second} for calibration, \SI{10}{\second} else & emulation time in biological time domain \\
  %%
  %\hline

  \hline\parameterhdr{3}{Other}{Software}\\\hline
  \bf Name & \bf Values & \bf Description\\\hline
  SpikeyHAL & 9e86d11c & git revision \\
  PyNN      & 2fe40b43 & git revision \\
  vmodule   & 76ef3b44 & git revision \\
  logger    & 826c5ed6 & git revision \\
  \hline\parameterhdr{3}{Other}{Hardware}\\\hline
  \bf Name & \bf Values & \bf Description\\\hline
  & chip 508 (version 5) & chip used for main manuscript \\
  & chip 503, 504 (version 5) and 603, 605, 666 (version 4) & chips used for supplements \\
  \hline
\end{tabularx}

\caption{
  Parameter values for the network model described in \prettyref{tab:nordlie}. 
  Bold numbers indicate default values.
  Gray numbers indicate target values not considering fixed-pattern noise.
  Leak conductances $\gleak$ are adjusted in the calibration process (see \prettyref{sub:methods_calib}).
  $\taumemeff$, $\tausyneff$, $\delayeff$ and $\vmaxeff$ describe effective values measured from spike-triggered averages in \rand{} emulations as described in \prettyref{app:fitting}.
  Effective values denote the median across synapses and trials followed by the 25th and 75th percentiles in brackets.
  \label{tab:nordlie_params}
}

\end{table*}

\endgroup
Details on the network model as well as parameter values are provided in \prettyref{tab:nordlie} and \ref{tab:nordlie_params}, respectively.
\section{Description of data analysis}
\begingroup
\squeezetable
\begin{table*}

\setlength{\columnwidthleft}{0.4\fullfigwidth}
\setlength{\columnwidthmiddle}{0.2\fullfigwidth}

\begin{tabularx}{\fullfigwidth}{|p{\columnwidthleft}|X|}
  \hline\modelhdr{2}{A}{Analysis measures}\\\hline
    \bf Measure                              & \bf Details \\
    \hline
    spike density                            & $\xi_i(t)=\sum_k \delta(t-t_i^k)$ \\
    \hline
    spike train                              & $\spiketrainindex(t_k)=$ number of spikes of neuron $i$ per bin $[k\tbin,(k+1)\tbin)$ \\
    \hline
    population activity                      & $\popactivity(t)=\frac{1}{N} \sum_i \spiketrainindex(t)$ \\
    \hline
    time-averaged population activity        & $\meanrate = \left<\popactivity(t)\right>_t$ \\
    \hline
    % \;                                     & with (finite time) Fourier transform \\
    % \;                                     & $S_i(f)=\mathfrak{F}[s_i(t)](f)=\int_0^T dt\, s_i(t)e^{-if t}$ \\
    membrane potential                       & $\memindex(t_k)=$ membrane potential of neuron $i$ in bin $[k\tbin,(k+1)\tbin)$\\
    \hline
    % \;                                     & with (finite time) Fourier transform \\
    (finite time) Fourier transform          & $X_i(f)=\mathfrak{F}[x_i(t)](f)=\int_0^T dt\, x_i(t)e^{-2\pi if t}$ (with inverse $\mathfrak{F}^{-1}$)\\
    \hline
    (single unit) power spectrum             & $\powerspec_i(f)=\frac{1}{T}X_i^*(f)X_i(f)$ \\
    \hline
    population-averaged power spectrum       & $\powerspec(f)=\frac{1}{N}\sum_i\powerspec_i(f)$ \\
    \hline
    population power spectrum                & $\poppowerspec(f)=\frac{1}{T}(\sum_iS_i^*(f))(\sum_jS_j(f))$ \\
    \hline
    pairwise cross spectrum                  & $\crossspec_{ij}=\frac{1}{T}X_i^*(f)X_j(f), i\ne j$ \\
    \hline
    population-averaged cross spectrum       & $\crossspec(f)=\frac{1}{N(N-1)}\sum_{i\ne j} \crossspec_{ij}(f)\equiv\frac{1}{\netsize(\netsize-1)}(\poppowerspec(f)-\netsize \powerspec(f))$ \\
                                             & (note: $\crossspec(f) \in \mathbb{R}$) \\
    \hline
    sliding window filter                    & $X(f) \rightarrow X(f) \ast H(f)$ \\
                                             & with $H(f)=\frac{1}{f_1-f_0}\Theta(f-f_0)\Theta(f_1-f)$ \\
    \hline
    coherence                                & $\lfc(f)=\frac{\crossspec(f)}{\powerspec(f)}$ \\
    \hline
    low-frequency coherence                  & $\lfc_X=\frac{1}{\fmax-\fmin}\int_{\fmin}^{\fmax} df \lfc(f)$ \\
    \hline
    pop.-averaged cross-correlation function & $\MakeLowercase{\crossspec}(\tau) = \frac{1}{\netsize(\netsize-1)}\sum_{i \ne j}\left<\sub{\spiketrain}{i}(t)\sub{\spiketrain}{j}(t+\tau)\right>_t \equiv \mathfrak{F}^{-1}[\crossspec(f)](\tau)$ \\
    \hline
    time average                             & $\left<\dots\right>_t$ \\
    \hline
  \end{tabularx}
  \caption{
    Summary of the data analysis. Here $i \in [1,\netsize]$, $X\in {S,V}$.
  }
  \label{tab:anasummary}
\end{table*}
\endgroup

\begingroup
\squeezetable
\begin{table*}

\setlength{\columnwidthleft}{0.15\fullfigwidth}
\setlength{\columnwidthmiddle}{0.55\fullfigwidth}

  \begin{tabularx}{\fullfigwidth}{| p{\columnwidthleft} | p{\columnwidthmiddle} | X |}
    \hline\parameterhdr{3}{A}{Analysis parameters}\\\hline
    \bf Parameter & \bf Description & \bf Values \\
    \hline
    % $\netsize$ & network size & [96, 112, 128, 144, 160, 176, \textbf{192}] \\
    $\tbin$ & bin size for spike trains & \SI{1}{\milli\second} \\
    \hline
    $\tbinmem$ & bin size for membrane potential traces & \SI{0.52}{\milli\second} \\
    \hline
    $\Twarmup$ & initial warm-up time (not considered in analysis) & \SI{1}{\second} \\
    \hline
    $\runtime$ & emulated network time (biol. time domain) & \SI{10}{\second} \\
    \hline
    $\Mtrials$ & number of network realizations & $\{50,\textbf{100}\}$ \\
    \hline
    $\Df$ & width of sliding window & \SI{1}{\hertz} \\
    \hline
    $\fmin,\fmax$ & interval boundaries for low-frequency coherence & \SI{0.1}{\hertz}, \SI{20}{\hertz} \\
    \hline
    $a$ & calibration state & $\{0, \frac{1}{8}, \frac{2}{8}, \frac{3}{8}, \frac{4}{8}, \frac{5}{8}, \frac{6}{8}, \frac{7}{8}, \textbf{1}\}$ \\
    \hline
  \end{tabularx}
  \caption{Summary of analysis parameters (default values in bold).}
  \label{tab:anaparams}
\end{table*}
\endgroup
A summary of the quantities and the data analysis used in this study is provided in \prettyref{tab:anasummary}. Parameter values for the data analysis are given in \prettyref{tab:anaparams}.

\bibliographystyle{neuralcomput_natbib_unsrt}
\bibliography{bib}

\ifarxivmode
\newif\ifarxivmode
\arxivmodetrue

\newcommand{\supp}{Supplements}

\ifarxivmode
\else
  \documentclass[reprint,aps,floatfix,a4paper,onecolumn]{revtex4-1}

\usepackage{stmaryrd}
\usepackage{amsfonts,amsmath}
\usepackage[T1]{fontenc}

\usepackage[utf8]{inputenc}
\usepackage{graphicx,prettyref,subfigure}
\usepackage[separate-uncertainty=true]{siunitx}
\usepackage{tabularx,color,colortbl} %tables
\usepackage{xr}
\usepackage{chngcntr}

\usepackage{enumitem} %compact lists
\setlist{noitemsep,topsep=0pt,parsep=0pt,partopsep=0pt}

\usepackage{setspace}
\usepackage{todonotes}

\newcommand{\smalltodo}[2][]
{\todo[size=\footnotesize, caption={#2}, #1]{\begin{spacing}{0.5}#2\end{spacing}}}
\newcommand{\smalltodored}[2][]
{\smalltodo[color=red, #1]{#2}}
\newcommand{\smalltodoyellow}[2][]
{\smalltodo[color=yellow, #1]{#2}}
\newcommand{\smalltodogreen}[2][]
{\smalltodo[color=green, #1]{#2}}
\newcommand{\verysmalltodo}[2][]
{\smalltodo[size=\tiny, #1]{#2}}

\renewcommand{\thesubfigure}{(\Alph{subfigure})}

\hyphenation{net-works semi-conduc-tor}

\newlength{\columnfigwidth}
\newlength{\fullfigwidth}
\setlength{\columnfigwidth}{8.6cm} %PRX
\setlength{\fullfigwidth}{17.2cm}

\newlength{\columnwidthleft}
\newlength{\columnwidthmiddle}

\newcommand{\modelhdr}[3]{
  \multicolumn{#1}{|l|}{
    \color{white}
    \cellcolor[gray]{0.0}
    \textbf{\makebox[0pt][l]{#2}\hspace{0.5\fullfigwidth}\makebox[0pt][c]{#3}}
  }
}
\newcommand{\parameterhdr}[3]{
  \multicolumn{#1}{|l|}{
    \color{black}\cellcolor[gray]{0.8}
    \textbf{\makebox[0pt][l]{#2}\hspace{0.5\fullfigwidth}\makebox[0pt][c]{#3}}
  }
}
  %\multicolumn{#1}{|l|}{
    %\color{black}\cellcolor[gray]{0.9}
    %\textit{\makebox[0pt]{#2}\hspace{0.5\fullfigwidth}\makebox[0pt][c]{#3}}
  %}

\renewcommand{\arraystretch}{1.5}

\newcommand{\mytitle}{The effect of heterogeneity on decorrelation mechanisms in spiking neural networks:\\ a neuromorphic-hardware study}
\newcommand{\mytitlenobreak}{The effect of heterogeneity on decorrelation mechanisms in spiking neural networks: a neuromorphic-hardware study}
\newcommand{\myauthor}{Pfeil et al.}

\usepackage[colorlinks=true,linkcolor=black,citecolor=black,urlcolor=black,filecolor=black,
pdftitle={\mytitlenobreak{}},pdfauthor={\myauthor{}},pdfsubject={},pdfkeywords={},
bookmarks=false,pdffitwindow]{hyperref}

\newrefformat{cap}{\hyperref[#1]{Figure~\ref{#1}}}
\newrefformat{fig}{\hyperref[#1]{Figure~\ref{#1}}}
\newrefformat{sec}{\hyperref[#1]{Section~\ref{#1}}}
\newrefformat{sub}{\hyperref[#1]{Section~\ref{#1}}}
\newrefformat{tab}{\hyperref[#1]{Table~\ref{#1}}}
\newrefformat{eq}{\hyperref[#1]{Equation~\ref{#1}}}
\newrefformat{app}{\hyperref[#1]{\ref{#1}}}
\newrefformat{appfig}{\hyperref[#1]{Supplements Figure~\ref{#1}}}
\newrefformat{apptab}{\hyperref[#1]{Supplements Table~\ref{#1}}}

\newcommand{\spikey}{\emph{Spikey}}                      % Spikey
\newcommand{\pynn}{\texttt{PyNN}}                        % PyNN
\newcommand{\fb}{FB}                                     % feedback network
\newcommand{\fbrep}{$\text{FB}_{\text{replay}}$}         % replayed network
\newcommand{\rand}{RAND}                                 % replayed network with input randomized
\newcommand{\mrmsub}[2]{#1_{\text{#2}}}                  % non-italic subscripts
\newcommand{\sub}[2]{#1_{#2}}                            % italic subscripts
\newcommand{\mrmsup}[2]{#1^{\text{#2}}}                  % non-italic superscripts

\newcommand{\ti}{t^*}                                    % indexed spike
\newcommand{\runtime}{T}                                 % runtime (biol. domain)
\newcommand{\exectime}{\mrmsub{\runtime}{exe}}           % time needed for execution of emulation (wall clock domain)
\renewcommand{\vec}[1]{{\bf #1}}
\newcommand{\mat}[1]{{\bf #1}}
\newcommand{\cvJ}{CV_J}                                  % coefficient of variation of synaptic weights

\newcommand{\netsize}{N}                                 % network size
\newcommand{\netsizeE}{N_\text{E}}                       % size of excitatory population
\newcommand{\netsizeI}{N_\text{I}}                       % size of inhibitory population
\newcommand{\noinputs}{K}                                % in-degree of each neuron
\newcommand{\noinputsE}{K_\text{E}}                      % excitatory in-degree of each neuron
\newcommand{\noinputsI}{K_\text{I}}                      % inhibitory in-degree of each neuron
\newcommand{\neuronindex}{i}                             % indexed neuron
\newcommand{\spikesource}{\xi}                           % off-chip spike source
\newcommand{\spikesourceindex}{\sub{\spikesource}{i}}    % indexed spike source

\newcommand{\cmem}{\mrmsub{C}{m}}                        % membrane capacitance
\newcommand{\vm}{v}                                      % membrane potential
\newcommand{\gleak}{\mrmsub{g}{l}}                       % leakage conductance
\newcommand{\gleaknull}{\mrmsub{g}{l,0}}                 % leakage conductance (uncalibrated)
\newcommand{\vrest}{\mrmsub{E}{l}}                       % resting potential
\newcommand{\vthresh}{\Theta}                            % firing threshold
\newcommand{\vreset}{\mrmsub{\vm}{reset}}                % reset potential
\newcommand{\vinh}{\mrmsub{E}{inh}}                      % inhibitory reversal potential
\newcommand{\tauref}{\mrmsub{\tau}{ref}}                 % refractory time
\newcommand{\taumem}{\mrmsub{\tau}{m}}                   % membrane time constant
\newcommand{\taumemeff}{\mrmsup{\taumem}{\text{eff}}}    % effective membrane time constant

\newcommand{\ginh}{\mrmsub{g}{syn}}                      % synaptic conductance
\newcommand{\gmax}{\mrmsub{g}{max}}                      % max conductance of hardware synapse
\newcommand{\tausyn}{\mrmsub{\tau}{syn}}                 % synaptic time constant
\newcommand{\tausyneff}{\mrmsup{\tausyn}{\text{eff}}}    % effective synaptic-current time constant
\newcommand{\weighthw}{\mrmsub{w}{hw}}                   % synaptic weight (in hardware stored as digital value)
\newcommand{\weight}{J}                                  % synaptic weight
\newcommand{\weightE}{J_\text{E}}                        % synaptic weight (exc)
\newcommand{\weightI}{J_\text{I}}                        % synaptic weight (inh)
\newcommand{\delay}{d}                                   % synaptic delay
\newcommand{\delayeff}{\mrmsup{d}{\text{eff}}}           % effective synaptic delay
\newcommand{\effweight}{w}                               % effective weight
\newcommand{\vmaxeff}{\mrmsup{\mrmsub{V}{max}}{\text{eff}}} % maximum effective post-synaptic potential

\newcommand{\rateindex}{\sub{r}{i}}                  % time average of firing rate for one neuron
\newcommand{\ratetime}{$\rate$(t)}                       % population average of firing rate for time step

\newcommand{\avrateneuroncalib}{\meanrate}               % target firing rate
\newcommand{\meanmem}{\bar{\vm}}                         % time average of membrane potential
\newcommand{\stdmem}{\sigma(\vm)}                        % standard deviation of above
\newcommand{\calibfactor}{a}                             % heterogeneity
\newcommand{\compfactor}{b}                              % calibration factor
\newcommand{\cvisi}{\mrmsub{CV}{ISI}}                 % coefficient of variation of interspike intervals
\newcommand{\dist}{D}                                    % distance of average membrane potential to threshold in standard deviations of membrane potential
\newcommand{\ratewidth}{\Delta r}                        % width of firing rate distribution

\newcommand{\spiketrain}{s}                              % spike train
\newcommand{\rate}{r}                                    % firing rate of neuron
\newcommand{\popactivity}{\bar{\spiketrain}}             % population activity
\newcommand{\meanrate}{\bar{\rate}}                      % time-averaged population activity
\newcommand{\spiketrainfourier}{\MakeUppercase{\spiketrain}} % Fourier transform of spike train
\newcommand{\spiketrainindex}{\sub{\spiketrain}{i}}      % spike train with index
\newcommand{\memindex}{\sub{\vm}{i}}                     % spike train with index
\newcommand{\memfourier}{\MakeUppercase{\vm}}            % Fourier transform of membrane potential
\newcommand{\powerspec}{A}                               % population-averaged power spectrum
\newcommand{\poppowerspec}{\bar{A}}                      % population power spectrum
\newcommand{\crossspec}{C}                               % population-averaged cross spectrum
\newcommand{\crossfunction}{\MakeLowercase{\crossspec}}  % population-averaged cross-correlation function
\newcommand{\lfc}{\kappa}                                % low-frequency coherence
\newcommand{\fmin}{\mrmsub{f}{min}}                      % integral boundries for LFC
\newcommand{\fmax}{\mrmsub{f}{max}}
\newcommand{\connectivity}{\epsilon}                     % connectivity
\newcommand{\tbin}{\Delta t}                             % bin size spike train
\newcommand{\tbinmem}{\Delta t_\mathrm{m}}               % bin size membrane potential
\newcommand{\Twarmup}{T_\text{warmup}}                   % burn-in time
\newcommand{\Mtrials}{M}                                 % number of network realizations
\newcommand{\Ltrials}{L}                                 % number of trials
\newcommand{\Df}{\Delta F}                               % sliding window to smoothen coherence

\newcommand{\taumax}{\tau_\mathrm{max}}                  % integration window for effective weights after trigger spikes
\newcommand{\taumin}{\tau_\mathrm{min}}                  % integration window for effective weights before trigger spikes

\renewcommand{\r}{\vec{r}}
\newcommand{\W}{\mat{W}}
\newcommand{\x}{\vec{x}}
\newcommand{\h}{h}
\newcommand{\diag}{\mathrm{diag}}
\newcommand{\CRR}{\mat{C}_\mathrm{RR}}
\newcommand{\CRRIN}{\mat{C}_\mathrm{RR}^\mathrm{in}}
\newcommand{\rff}{\tilde{\r}}
\newcommand{\q}{\vec{q}}
\newcommand{\CRRFF}{\mat{C}_\mathrm{\tilde{R}\tilde{R}}}
\newcommand{\CQQIN}{\mat{C}_\mathrm{QQ}^\mathrm{in}}
\newcommand{\CQQ}{\mat{C}_\mathrm{QQ}}
\newcommand{\ABARX}{\bar{A}_\mathrm{X}}
\newcommand{\CXXii}{C_{\mathrm{XX},ii}}
\newcommand{\CBARXX}{\bar{C}_\mathrm{XX}}
\newcommand{\CXXij}{C_{\mathrm{XX},ij}}
\newcommand{\muw}{\mu_\mathrm{\effweight}}
\newcommand{\sigmaw}{\sigma_\mathrm{\effweight}}
\newcommand{\muW}{\mu_\mathrm{\MakeUppercase{\effweight}}}
\newcommand{\sigmaW}{\sigma_\mathrm{\MakeUppercase{\effweight}}}

\newcommand{\dither}{\vartheta}
\newcommand{\stdweight}{\sigma_\weight}
\newcommand{\stdthresh}{\sigma_\vthresh}
\newcommand{\aveffweight}{\bar{\effweight}}

\newcommand{\Isyn}{\mrmsub{I}{syn}}
\newcommand{\numrecordspikesE}{P_\mathrm{\spiketrain}^\text{E}}
\newcommand{\numrecordmemE}{P_\mathrm{\vm}^\text{E}}
\newcommand{\numrecordspikesI}{P_\mathrm{\spiketrain}^\text{I}}
\newcommand{\numrecordmemI}{P_\mathrm{\vm}^\text{I}}
\newcommand{\numrecordspikes}{P_\mathrm{\spiketrain}}
\newcommand{\numrecordmem}{P_\mathrm{\vm}}

\definecolor{graytarget}{gray}{0.5}
  \externaldocument{main}
\fi

\renewcommand{\figurename}{SUP. FIG.}
\renewcommand{\tablename}{SUP. TABLE}

\ifarxivmode
  \widetext
  \clearpage
  \begin{center}
    \textbf{\large \supp{}}
  \end{center}

  \setcounter{section}{0}
  \setcounter{equation}{0}
  \setcounter{figure}{0}
  \setcounter{table}{0}
  \setcounter{page}{1}

  \renewcommand{\appendixname}{}
  \renewcommand{\appendixesname}{}

  \counterwithout{equation}{section}
\else
  \begin{document}

  \begin{center}
    \textbf{\large \supp{}}
  \end{center}
\fi

\renewcommand{\theequation}{S\arabic{equation}}

\pagenumbering{roman}

\renewcommand{\thesection}{Supplements \arabic{section}}

\section{Power consumption}
\label{app:power}

The \spikey{} system consumes approximately \SI{6}{\watt} of power, and the chip itself less than \SI{0.6}{\watt}.
On the chip most power is consumed by digital communication infrastructure, which is not part of the neuromorphic network.
In the following, we estimate the power consumption for a single synaptic event using the data set partly shown in \prettyref{fig:raster}a.
This emulation lasts $\runtime = \SI{10}{\second}$ in biological time and generates approximately $45 \cdot 10^3$ spikes.
Considering the acceleration of the hardware network ($10^4$) and the synapse count per neuron ($\noinputs=15$), the system generates $7 \cdot 10^8$ synaptic events per second in hardware time.
If we consider the total power consumption of the \spikey{} chip, the upper bound of energy consumed by each synaptic transmission will be approximately \SI{1}{\nano\joule}.
Because these measurements include the communication infrastructure and other support electronics to observe spike times and membrane traces,
the real energy consumption for synaptic transmissions is estimated to be approximately ten times smaller.
Network simulations on conventional supercomputers a far less energy efficient and consume tens of \SI{}{\micro\joule} for each synaptic transmission \citep{Sharp12}.

\section{Modification of the bisection method}
\label{app:binary_noisy}

In each iteration of the bisection method that is used to calibrate the leak conductances of hardware neurons (\prettyref{sub:methods_calib}),
we evaluated the firing rate for each neuron by the median over $\Ltrials=25$ identical trials.
However, if this measure is compared between consecutive identical iterations, temporal noise on time scales longer than the duration of one iteration may still lead to variability.
In the original bisection method, the interval of possible solutions is halved after each iteration step \citep{Press07}.
To improve the convergence of this method in the context of our calibration we expanded the halved interval by $20\%$ at both ends after each iteration.
This prevents the algorithm to get stuck in an interval wrongly chosen by random fluctuations of the firing rates.

\section{Effective weights}
\label{app:weff}
\begin{figure*}[!b]
\centering
\includegraphics{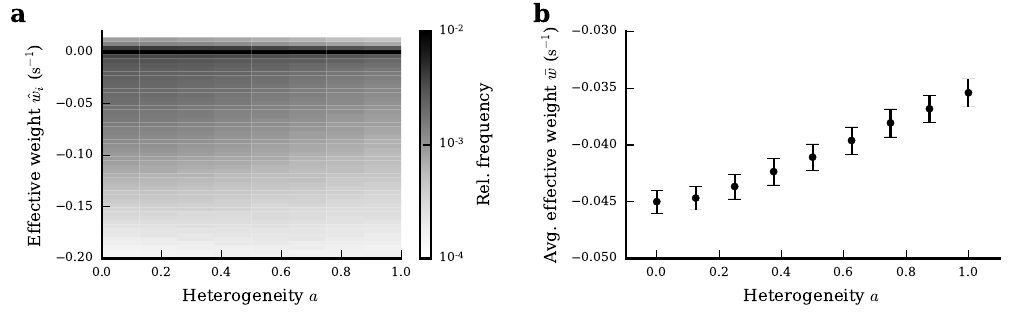}
\caption{
  Distribution of effective weights $\effweight$ (a; log scale) and population-averaged effective weights $\aveffweight$ (b) for different levels of heterogeneity, averaged over $\Mtrials=100$ network realizations.
  Symbols and error bars denote mean and one standard deviation, respectively.
  \label{appfig:weff}
}
\end{figure*}

We quantify the effect of a single spike of neuron $j$ on the firing rate of a postsynaptic neuron $i$ by the effective weight $\effweight_{ij}$ of the connection $i \leftarrow j$.
Assuming that the activity of neuron $i$ does not affect the activity of neuron $j$ (i.e., the \rand{} case), we define $\effweight_{ij}$ as the cross-correlation between the spike trains $\spiketrain_j(t)$ and $\spiketrain_i(t + \tau)$, where, due to causality, $\tau$ is positive.
Then, we average the effective weight over the emulated network time $\runtime$, and subtract the baseline determined by the average correlation for negative $\tau$:
\begin{align}
  \effweight_{ij} =& \frac{1}{\taumax} \int_0 ^{\taumax} \langle \spiketrain_j(t) \spiketrain_i(t+\tau) \rangle_t d\tau - \frac{1}{\taumin} \int_{-\taumin} ^{0} \langle \spiketrain_j(t) \spiketrain_i(t+\tau) \rangle_t d\tau \quad .
\end{align}
Here, we chose $\taumax=\SI{50}{\milli\second}$ and $\taumin=\SI{50}{\milli\second}$.
$\taumax$ was determined by measuring the average duration in which a spike from neuron $j$ has an influence on neuron $i$ (data not shown).
$\taumin$ was then chosen symmetrically.
The mass of the effective weights density shifts towards less negative effective weights for increasing heterogeneity (\prettyref{appfig:weff}a).

\begin{figure*}
\centering
\includegraphics{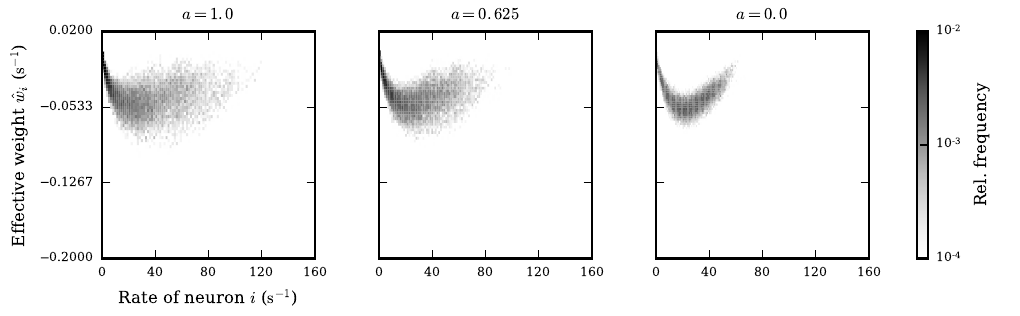}
\caption{
  Distribution of average incoming effective weights $\hat{\effweight}_i$ over the rates of the postsynaptic neurons $i$, for three different levels of heterogeneity ($a=0$, $a=0.625$ and $a=1$), averaged over $\Mtrials=100$ network realizations.
  The average incoming effective weights are obtained by summing over all presynaptic partners ($\hat{\effweight}_i=\sum_j \effweight_{ij}$).
  \label{appfig:weff_ratehist}
}
\end{figure*}

We obtain $\bar{\effweight}$ by averaging over all possible connections:
\begin{align}
  \aveffweight = \frac{1}{\netsize\noinputs} \sum_{i,j} \effweight_{ij} \quad .
\end{align}
For increasing heterogeneity the absolute value of the average effective weight $\aveffweight$ decreases (\prettyref{appfig:weff}b).
This can be explained by the dependence of the effective weight on the firing rate of the postsynaptic neuron (\prettyref{appfig:weff_ratehist}).
In the regime of small rates ($<\SI{10}{\second}^{-1}$), incoming spikes hardly affect the neuron's firing and the effective weight is small.
Similarly, in the case for large rates ($>\SI{40}{\second}^{-1}$).
Neurons with intermediate firing rates are sensitive to input and hence have a more negative effective weight.
Heterogeneity increases the number of neurons with small and large firing rates, and hence the absolute value of the average effective weight decreases, which in turn weakens the effective negative feedback of the network.

\section{Simulations with software}
\label{app:sim}

\begin{figure*}
\centering
\includegraphics{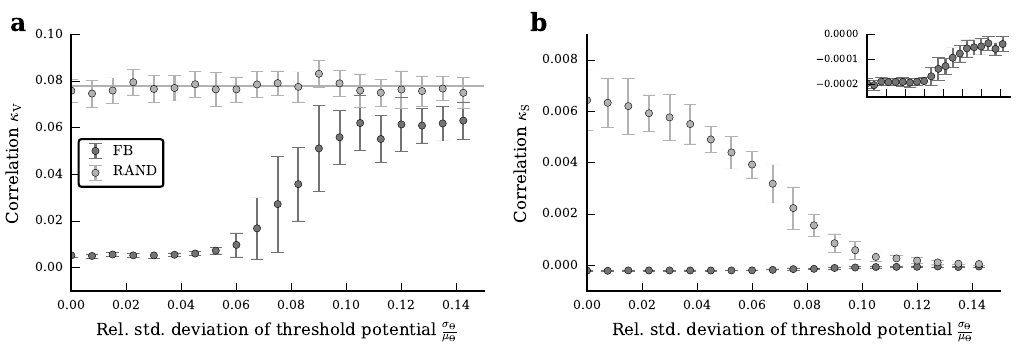} % N=25*192=4800, K=375

\caption{
  Like \prettyref{fig:sim}, but for $\netsize=4800$ neurons and $\noinputs=375$ inputs for each neuron.
  Note the different scales of the ordinate in \prettyref{fig:sim} and (b).
  Symbols and error bars denote mean and standard deviation, respectively, across $\Mtrials=20$ network realizations (error bars are partly covered by markers).
  % The gray line in (a) depicts shared-input correlations in a homogeneous network (\prettyref{eq:shared_input}).
  % The inset in (b) shows a magnified view of the spike-train correlations in the \fb{} case.
  % Note that in simulations the \fbrep{} is identical to the \fb{} case, and is hence not shown.
  % Networks simulated with NEST \citep{Gewaltig_07_11204} and PyNN \citep{Davison08_11}.
  \label{appfig:sim}
}
\end{figure*}
We validate our results by comparing them to simulations with software (NEST \citep{Gewaltig_07_11204}, PyNN \citep{Davison08_11}).
In these simulations we modulated the degree of heterogeneity by distributing the firing thresholds of all neurons according to a normal distribution with mean $\vthresh$ and standard deviation $\stdthresh$.
Details about the network, neuron and synapse models and their parameters can be found in \prettyref{apptab:sim_nordlie} and \ref{apptab:sim_nordlie_params}, respectively.
The results are qualitatively similar to network emulations on the \spikey{} chip (compare \prettyref{fig:corr_calib} to \ref{fig:sim}) and also hold for larger network sizes (\prettyref{appfig:sim}).
In the \fb{} case, input correlations and output correlations increase with network heterogeneity.
In contrast to hardware emulations, input correlations stay approximately constant in the \rand{} case.
Output correlations strongly decrease with the standard deviation $\stdthresh$.
Here, heterogeneity only affects the output spike times, and not the integrative properties of the neurons (see also \prettyref{sub:results_heterogeneity} and \citep{Yim2013_032710}).

\begin{figure*}
\centering
\includegraphics{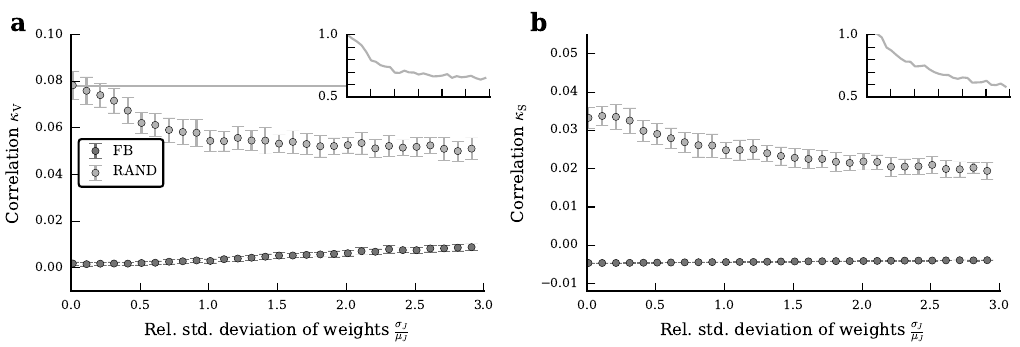}

\caption{
  Dependence of population-averaged input correlations \textbf{(a)}, and spike-train correlations \textbf{(b)} on the width of the weight distribution, for the intact network (\fb{}, dark gray circles) and the \rand{} (light gray circles) case.
  Symbols and error bars denote mean and standard deviation, respectively, across $\Mtrials=30$ network realizations (error bars are partly covered by markers).
  The gray line in (a) depicts shared-input correlations in a homogeneous network (\prettyref{eq:shared_input}).
  The insets show correlations normalized to unity at $\stdweight=0$, with the same abscissa as in the main plot.
  Note that in simulations the \fbrep{} is identical to the \fb{} case, and is hence not shown.
  Networks simulated with NEST \citep{Gewaltig_07_11204} and PyNN \citep{Davison08_11}.
}
\label{appfig:sim_weights}
\end{figure*}
In addition, we compare the effect of heterogeneity in firing thresholds on correlations to that of heterogeneity in synaptic weights.
Details about the network, neuron and synapse models and their parameters can be found in \prettyref{apptab:sim_nordlie} and \ref{apptab:sim_nordlie_params_weights}, respectively.
If we distribute synaptic weights, shared-input correlations decrease with the width of the weight distribution (\prettyref{appfig:sim_weights}).
Output correlations are also reduced, however, only proportionally to the reduction of input correlations (compare insets in \prettyref{appfig:sim_weights}).
While output correlations are overall smaller than input correlations due to the non-linearity in spike generation, we do not observe a boost of this decrease for large heterogeneities.
Overall, the dynamics of the recurrent system is more sensitive to heterogeneities in firing thresholds than in synaptic weights (compare scale of abscissas of \prettyref{fig:sim} to \prettyref{appfig:sim_weights}).

\begin{figure*}
\centering
\includegraphics{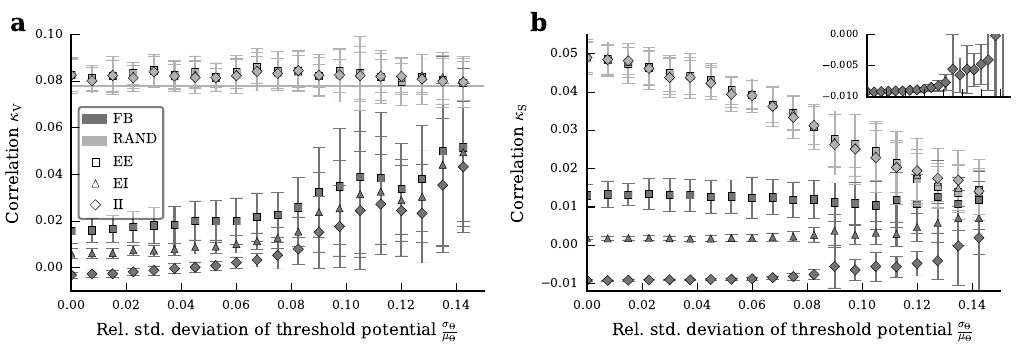} % N=192; E/I randomized

\caption{
  Dependence of population-averaged input correlations \textbf{(a)}, and spike-train correlations \textbf{(b)} on the width of the threshold distribution, for the intact network (\fb{}, dark gray circles) and the \rand{} (light gray circles) case.
  The network consists of two populations, one excitatory and one inhibitory, of equal size.
  Symbols and error bars denote mean and standard deviation, respectively, across $\Mtrials=30$ network realizations (error bars are partly covered by markers).
  The gray line in (a) depicts shared-input correlations in a homogeneous network (\prettyref{eq:shared_input}).
  The inset in (b) shows a magnified view of the spike-train correlations in the \fb{} case.
  Note that in simulations the \fbrep{} is identical to the \fb{} case, and is hence not shown.
  Networks are simulated with NEST \citep{Gewaltig_07_11204} and PyNN \citep{Davison08_11}.
}
\label{appfig:sim_ei}
\end{figure*}
To investigate the generalization of our results to mixed excitatory-inhibitory networks, correlations were measured in a network of $\netsize=192$ neurons consisting of half excitatory and half inhibitory neurons.
Details about the network, neuron and synapse models and their parameters can be found in \prettyref{apptab:sim_nordlie_EI} and \ref{apptab:sim_nordlie_params_EI}, respectively.
As described above, we distribute the firing thresholds of all neurons according to a normal distribution.
The results are consistent with those obtained from purely inhibitory networks on hardware demonstrating the generality of our findings (compare \prettyref{fig:corr_calib} to \prettyref{appfig:sim_ei}).
By modulating the level of heterogeneity separately for the excitatory or the inhibitory population, we observe that network dynamics are more sensitive to heterogeneities in the inhibitory than in the excitatory population (data not shown).

\begin{figure*}
\centering
\includegraphics{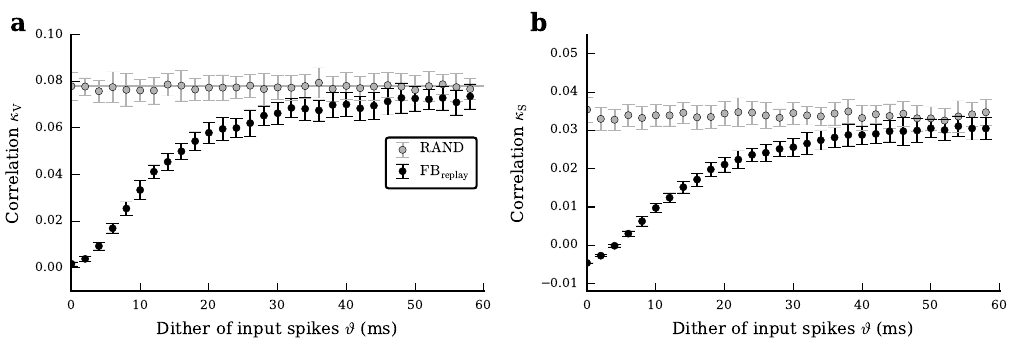}

\caption{
  Dependence of population-averaged input correlations \textbf{(a)}, and spike-train correlations \textbf{(b)} on the strength of temporal noise, for the \fbrep{} (black circles) and the \rand{} (light gray circles) case.
  Symbols and error bars denote mean and standard deviation, respectively, across $\Mtrials=30$ network realizations (error bars are partly covered by markers).
  Gray curve in (a) depicts shared-input correlations in a homogeneous network (\prettyref{eq:shared_input}).
  % The light gray horizontal band represents mean $\pm$ three standard deviations of correlations in surrogate data, in which spatial correlations were removed.
  Correlations in the \fb{} case correspond to correlations in the \fbrep{} case for $\dither=\SI{0}{\milli\second}$.
  Networks simulated with NEST \citep{Gewaltig_07_11204} and PyNN \citep{Davison08_11}.
}
\label{appfig:sim_dither}
\end{figure*}

\begin{figure*}
\centering
\includegraphics{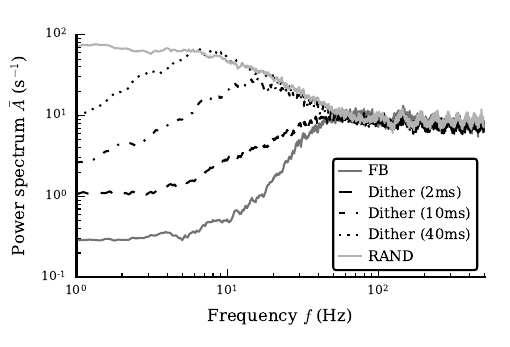}

\caption{Power spectrum of the population activity of the networks in
  \prettyref{appfig:sim_dither} for the intact network (solid dark gray),
  different degrees of dither on input spikes (dashed, dashed dotted and dotted black) and
  the \rand{} case (solid light gray), averaged across $\Mtrials=30$ network realizations.}
\label{appfig:sim_dither_power}
\end{figure*}
We investigate the effect of temporal noise on correlations by dithering spikes in the \fbrep{} case before replaying them to the network (for network, neuron and synapse models see \prettyref{apptab:sim_nordlie} and \ref{apptab:sim_nordlie_params}).
Each spike time $t$ is replaced by a spike time $t'$ randomly drawn from a normal distribution with width $\dither$: $t' \sim \mathcal{N}(t, \dither)$.
Even for small $\dither$, correlations increase on the input as well as on the output side, demonstrating the sensitivity of correlations to perturbations in the feedback loop (see \prettyref{appfig:sim_dither} and \cite{Tetzlaff12_e1002596}).
This effect is also reflected in an increase of the power of the population activity at small frequencies (see \prettyref{appfig:sim_dither_power}).
These results suggest that on hardware temporal noise is responsible for the increase of correlations and population power in the \fbrep{} case (compare \fbrep{} to \fb{} in \prettyref{fig:raster}, \ref{fig:corr_n} and \ref{fig:corr_calib}).

\section{Linear model}
\label{app:linear}

We investigate the consistency of our results with a linear rate model that allows us to numerically calculate the average correlations from a given connectivity matrix $\W$.
The model is defined as (according to, e.g., \citep{Grytskyy13_131})
\begin{align}
  \r(t) &= (\W(\r+\x)\ast \h)(t) \quad .
  \label{eq:app_linear}
\end{align}
Here, $\r(t)$ denotes the rate of the individual neurons and $\x(t)$ a Gaussian white noise input that is independent for each neuron.
The linear filter kernel $\h(t)$ depends on the details of the model, is not relevant in our calculation, and hence is not further specified here.
\prettyref{eq:app_linear} can be transformed to Fourier domain, where the input and output spectral matrices can be expressed by
\begin{align}
  \CRR(\omega) &= T(\omega)T(\omega)^\dagger \quad ,
  \label{eq:app_lincoout} \\
  \CRRIN(\omega) &= \W\CRR(\omega)\W^T \quad ,
  \label{eq:app_lincoin}
\end{align}
with $T(\omega)=(1-H(\omega) \W)^{-1}$ \citep{Grytskyy13_131}.
In the \rand{} case, the linear equation for the rate of the (unconnected) neurons reads
\begin{align}
  \q(t) = (\W(\rff+\x)\ast \h)(t) \quad ,
  % \label{eq:app_linearff}
\end{align}
where $\rff(t)$ has the same auto-correlations as $\r(t)$ but zero cross correlations, i.e., $\CRRFF = \diag(\CRR)$, since the randomization of spike times removes all spatio-temporal correlations.
According to \citet{Tetzlaff12_e1002596} spectral matrices in the \rand{} case are given by
\begin{align}
  \CQQIN(\omega) =& \W \CRRFF \W^T \quad ,
  \label{eq:app_lincoinff} \\
  \CQQ(\omega) =& |H(\omega)|^2(\CQQIN + \rho) \quad .
  \label{eq:app_lincooutff}
\end{align}
We calculate the population-averaged power- and cross-spectra from the full matrices:
\begin{align}
  \ABARX(\omega) &= \frac{1}{\netsize}\sum_i\CXXii \quad , \\
  \CBARXX(\omega) &= \frac{1}{\netsize(\netsize-1)}\sum_{i\ne j}\CXXij \quad .
\end{align}
Here, $X\in \{R,Q\}$ denotes the \fb{} and \rand{} case, respectively.
The low frequency coherence is the cross-spectra normalized by the power spectra:
\begin{align}
  \kappa_X(0)=\frac{\CBARXX(0)}{\ABARX(0)} \quad .
\end{align}
Note that in the linear model we are actually taking the zero frequency coherence.

\begin{figure*}
\centering
\includegraphics{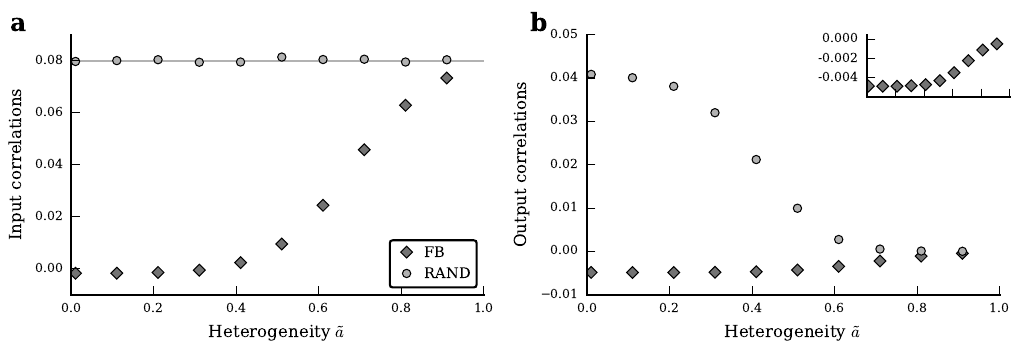}
\caption{
  Dependence of population-averaged input correlations \textbf{(a)}, and output correlations \textbf{(b)} on the absolute magnitude of the effective weights (``heterogeneity'' $\tilde{a}$, see text), for the intact network (\fb{}, dark gray diamonds) and the \rand{} (light gray circles) case.
  Gray curve in (a) depicts shared-input correlations in a homogeneous network (\prettyref{eq:shared_input}).
  The inset in (b) shows a magnified view of the spike-train correlations in the \fb{} case (dark gray diamonds).
  Note that in the linear model the \fbrep{} is identical to the \fb{} case, and is hence not shown.
  \label{appfig:linear_corr}
}
\end{figure*}

As in the spiking model, we consider a sparse network, i.e., we randomly choose for each neuron $i \in [1,\netsize]$ an identical number of presynaptic partners ($\noinputs=15$).
In the linear model we do not consider a distribution of non-zero effective weights.
Instead, each realized connection is assigned the same weight value $-\effweight$.
To mimic the effect of calibration we vary the absolute value of the effective weight by scaling the weights of the non-zero connections with a sigmoidal function of $\tilde{\calibfactor}\in [0,1]$:
\begin{align}
  \tilde{\effweight}=\frac{1}{1+e^{10\times(\tilde{\calibfactor}-0.5)}}\effweight \quad .
\end{align}
This procedure changes the variance of the weight matrix \cite{Bernacchia13_1732} and hence $\tilde{\calibfactor}$ is denoted the \emph{heterogeneity} of the network.
More homogeneous (heterogeneous) networks have larger (smaller) effective weights and hence stronger (weaker) feedback.
We obtain qualitatively similar results as we observe on the \spikey{} chip (compare \prettyref{appfig:linear_corr} to \prettyref{fig:corr_calib}).
Output correlations in the \rand{} case decrease, while both input and output correlations in the \fb{} case increase with network heterogeneity, i.e., with the variance of the effective weight matrix.
Note that, in this simplified model, input correlations in the \rand{} case are not affected by the level of "heterogeneity" because the non-zero weights are homogeneous, i.e.~have zero variance, irrespectively of $\tilde{\calibfactor}$ (cf.~\prettyref{eq:shared_input_hetero}).

\begin{figure}
  \includegraphics{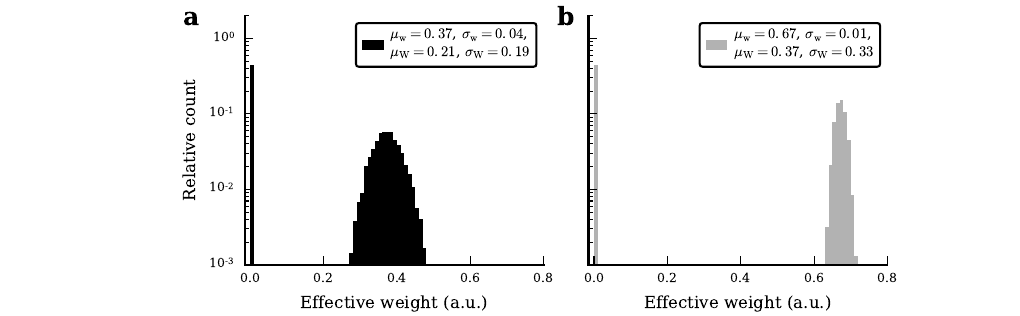}
  \caption{
    Distributions of effective weights in a sparse network.
    Different values for the mean $\muw$ and standard deviation $\sigmaw$ of the distribution of non-zero weights in \textbf{(a)} and \textbf{(b)}, respectively,
    result in different values for the mean $\muW$ and standard deviation $\sigmaW$ of the effective weight matrix.
  }
  \label{appfig:effective_weight_dist}
\end{figure}

In \prettyref{appfig:effective_weight_dist} we illustrate, how in a sparse network the variance of the weight matrix can increase, although the distribution of non-zero weights becomes narrower.
The standard deviation $\sigmaw$ of a distribution of non-zero weights with mean $\muw$ is (in this example) smaller than the standard deviation $\sigmaW$ of the full effective weight matrix,
due to the sparseness of the matrix (\prettyref{appfig:effective_weight_dist}a; here, we chose $\connectivity=0.8$).
If we, at the same time, increase the mean $\muw$ and decrease the standard deviation $\sigmaw$ of non-zero weights,
the standard deviation of the weight matrix $\sigmaW$ can increase significantly (\prettyref{appfig:effective_weight_dist}).
While the distribution of effective weights is broadened, the mean is decreased, which has a greater impact on the size of correlations.
This observation could explain the decrease of correlations with increased calibration of the neuromorphic chip.

\section{Firing statistics in the \rand{} case}
\label{app:statistics_rand}
\begin{figure*}
\includegraphics{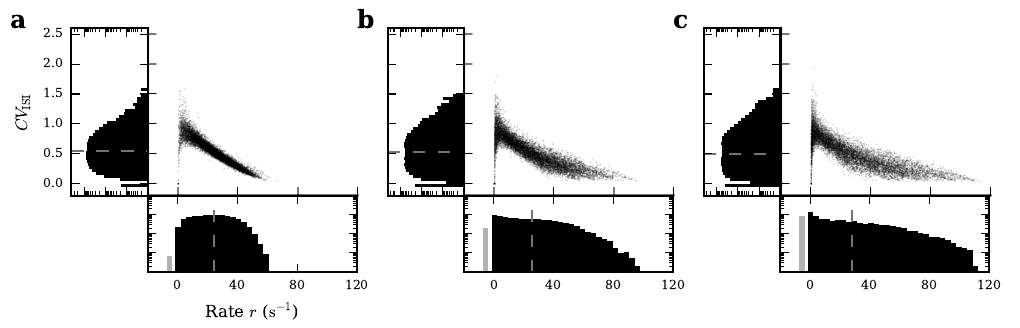}
\caption{
\textbf{(a--c)}
Like \prettyref{fig:rates_cv}d--f, but for the \rand{} case.
The mean of firing rate (\SI{24.0}{\per\second}, \SI{25.2}{\per\second}, \SI{27.8}{\per\second}) and $\cvisi$ distributions ($0.54$, $0.53$, $0.5$) are marked with dashed lines.
Percentage of dead neurons: $0.00\%, 0.02\%, 0.08\%$.
\label{appfig:rates_cv_ff}
}
\end{figure*}

The firing rate distributions in the \rand{} case are similar to those in the \fb{} scenario (compare \prettyref{appfig:rates_cv_ff} to \prettyref{fig:rates_cv}d--f).
As in the \fb{} case they become broader for increasing heterogeneity.
Due to a larger variance of the membrane potentials, inactive neurons in the \fb{} case have a higher probability to fire in the \rand{} case (compare gray bars in horizontal histograms of \prettyref{appfig:rates_cv_ff}a--c and \prettyref{fig:rates_cv}d--f).
Neurons with firing rates larger than the average firing rate are strongly driven by constant current influx, and hence show similar firing rates than in the \fb{} scenario.
The irregularity of firing increases for the \rand{} compared to the \fb{} case.
This can also be traced back to an increased variance of the membrane potential for the \rand{} case (data not shown).

\section{Correlation matrices}
\label{app:corrmatrix}
\begin{figure}
\includegraphics{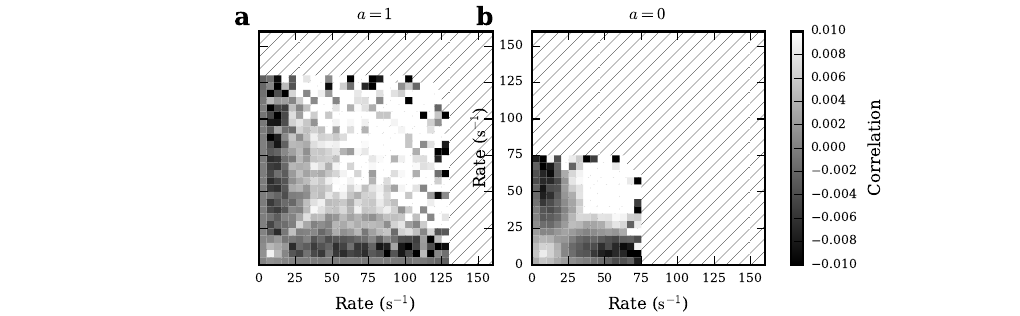}
\caption{
  %Pairwise spike-train correlations for all pairs of neurons in $\Mtrials=100$ network realizations, sorted by the average firing rate of the respective neurons.
  %Diagonal stripes mark regions where no data was available.
  %Here $C_{ij}(f) \in \mathbb{C}$, but we only consider the real part of the cross spectrum to calculate the low-frequency coherence: $\lfc_{ij}=\frac{1}{\fmax-\fmin}\int_{\fmin}^{\fmax} \frac{\mathfrak{Re}(C_{ij}(f))}{\sqrt{A_i(f) A_j(f)}}$.
  Firing-rate dependence of pairwise low-frequency spike-train coherence in the uncalibrated {\bf (a)} and fully calibrated system {\bf (b)}.
  Each bin represents the low-frequency spike-train coherence $\lfc_{ij}=\frac{1}{\fmax-\fmin}\int_{\fmin}^{\fmax} \frac{\mathfrak{Re}(C_{ij}(f))}{\sqrt{A_i(f) A_j(f)}}$, averaged across all neuron pairs $\{i,j\}$ with firing rates in the interval $[r_i,r_i+\Delta{}r)$ and $[r_j,r_j+\Delta{}r)$. 
  Here, $\mathfrak{Re}(C_{ij}(f))$ denotes the real part of the complex cross-spectrum at frequency $f$.
  $\fmin=\SI{0.1}{\hertz}$, $\fmax=\SI{20}{\hertz}$, $\Delta{}r=\SI{5}{\second}^{-1}$.
  Data averaged across $\Mtrials=100$ network realizations. 
}
\label{appfig:corrmatrix}
\end{figure}

In addition to the population-averaged measures from the main manuscript,
we calculated the pairwise correlations for each pair $i,j$ of neurons with $i\ne j \in [1, \netsize]$,
and ordered these by the time-averaged rate of the corresponding neurons (\prettyref{appfig:corrmatrix}).
This reveals a dependence of the pairwise correlation on the firing rate of the respective neurons.
If both neurons fire at low rate (here $<\SI{5}{\per\second}$), correlations will be close to zero similar to the results in \cite{DeLaRocha07_802}.
For high rates (here $>\SI{25}{\per\second}$) we find mostly positive correlations.
However, for neurons firing at intermediate rates, the activity of the pair is often anti-correlated,
and can hence effectively suppress (positive) shared-input correlations.
After calibration, the amount of neurons firing at intermediate rates and showing negative correlations increases (\prettyref{appfig:corrmatrix}b, \prettyref{fig:rates_cv}).
Shared-input correlations are hence suppressed by more neurons leading to smaller input correlations than in the uncalibrated case (\prettyref{fig:corr_calib}).

\section{Results for different \spikey{} chips}
\label{app:chips}

\begin{figure*}
\centering
\includegraphics{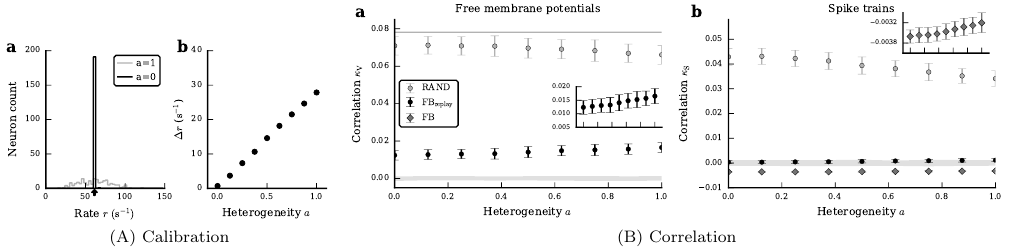}
\includegraphics{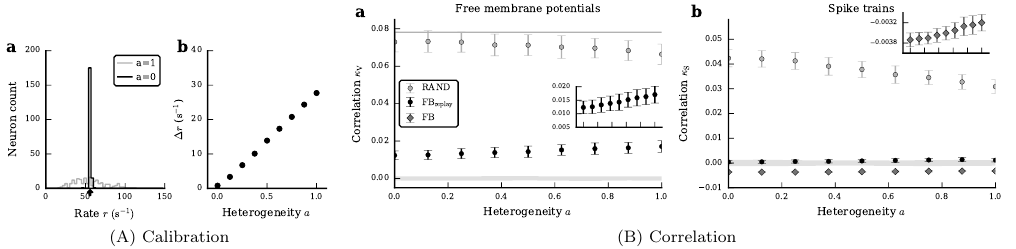}
\caption{
Like \prettyref{fig:calib} and \ref{fig:corr_calib} in (A) and (B), respectively, but for two additional chips of version 5.
}
\label{appfig:chipsv5}
\end{figure*}

We have performed the same experiments as described in the main text using two additional \spikey{} chips of the current version (5).
Different chips show different realizations of fixed-pattern noise, and hence calibration was repeated for each chip separately (\prettyref{fig:calib}, \prettyref{appfig:chipsv5}A).
For all chips we find qualitatively the same results as described above: in the \fb{} case, free membrane potentials (and spike trains) are decorrelated by inhibitory feedback and correlations increase with an increasing level of heterogeneity (\prettyref{fig:corr_calib}, \prettyref{appfig:chipsv5}B).

\begin{figure*}
\centering
\includegraphics{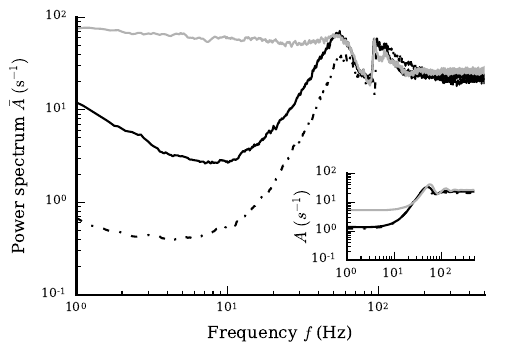}
\includegraphics{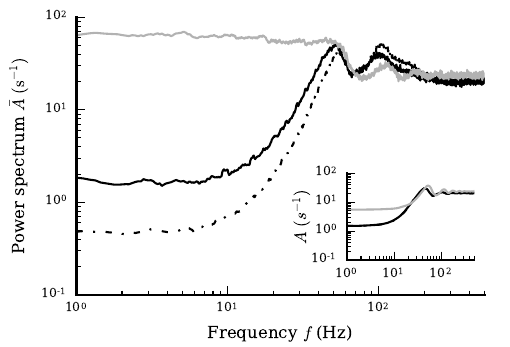}
\caption{
  Spectra of trial-averaged population power $\poppowerspec(f)$ and trial-averaged single-cell power $\powerspec(f)$ for chip 666 (version 4, left) and chip 508 (version 5, right) illustrating the synchronization artifact in the \SI{100}{\hertz} range for the intact network (\fb{}, dark gray), the \fbrep{} (black) and the \rand{} (light gray) case.
}
\label{appfig:powerspec_raster}
\end{figure*}
\begin{figure*}
\centering
\includegraphics{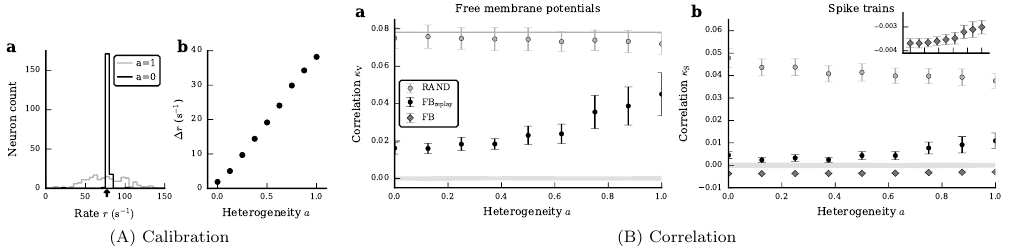}
\includegraphics{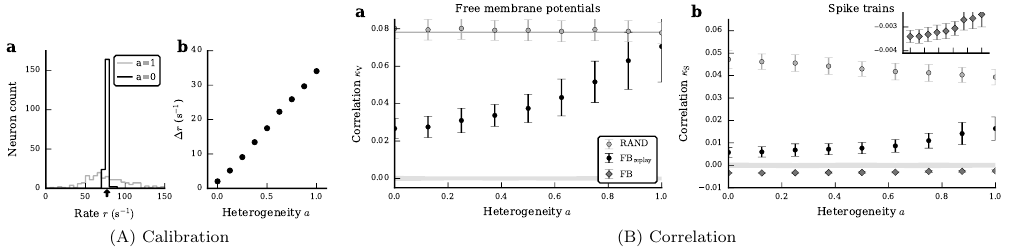}
\includegraphics{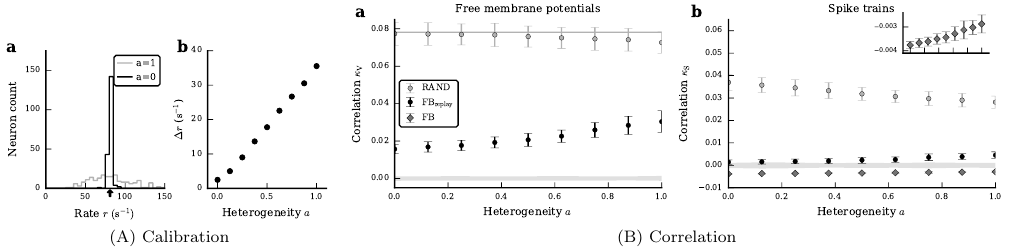}
\caption{
Like \prettyref{fig:calib} and \ref{fig:corr_calib} in (A) and (B), respectively, but for three chips of version 4.
}
\label{appfig:chipsv4}
\end{figure*}

This also applies to the results of identical experiments on three previous-generation chips (version 4).
These old chips show more pronounced heterogeneity in their parameters.
In addition, we observe larger quantitative differences between old chips for the uncalibrated case, which is likely to be caused by different extents of intrinsic fixed-pattern noise.
Unfortunately, the old systems show an artifact affecting the population power at around \SI{100}{\hertz} (\prettyref{appfig:powerspec_raster}), consistently across all chips of identical revision.
We can exclude that this effect can be traced back to network effects (it occurs across the \fb{}, \fbrep{} and \rand{} cases) or to single-cell power spectra (data not shown).
We hence have to consider it an artificial spurious synchronization.
As the artifact lies outside the range of frequencies that are relevant for our measures of correlation it does not affect the main results and experiments on the old chips qualitatively confirm our results in the main manuscript (compare \prettyref{appfig:chipsv4} to \prettyref{fig:calib}, \ref{fig:corr_calib} and \prettyref{appfig:chipsv5}).

\section{Reproducibility of networks with intact feedback}
\label{app:control}
\begin{figure}
\includegraphics{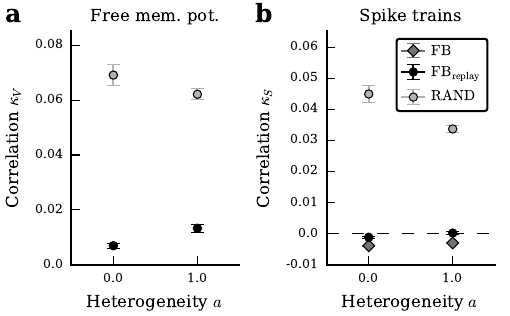}
\caption{
  Like \prettyref{fig:corr_calib} and for the same chip, but for $\Ltrials=20$ trials of one network realization.
  }
\label{appfig:control_calib}
\end{figure}

We measured the standard deviation of free membrane potential and spike train correlations over several trials for a single network realization (\prettyref{appfig:control_calib}).
This standard deviation is smaller than the standard deviation we observe over different network realizations (compare to \prettyref{fig:corr_calib} and \prettyref{appfig:chipsv5}),
which indicates that the latter is mostly caused by the different realizations of the connectivity, not by trial-to-trial variability.
Note that the variability between trials of networks with intact feedback is likely to be larger than between replays of network activity as shown in \prettyref{fig:reprod}, because network dynamics may be chaotic.
The data shown in \prettyref{appfig:control_calib} has to be interpreted with care, because the reproducibility of only a single network realization is considered.
For different realizations, the mean and standard deviation will in general be different.

\section{Extraction of effective PSP amplitudes, time constants and delays}
\label{app:fitting}

In order to provide an estimation for the order of magnitude of various neuron parameters, we fitted the postsynaptic potentials (PSPs) of current-based LIF neurons with exponential postsynaptic currents (PSCs) to the spike-triggered average (STA) of free membrane potentials obtained from hardware emulations.
The differential equations describing the dynamics of membrane potentials $V_j(t)$ and synaptic currents $I_j(t)$ for this simplified model are given by
\begin{align}
  \taumemeff \dot{V}_j(t) =& -V_j(t) + R I_j(t) \\
  \tausyneff \dot{I}_j(t) =& -I_j(t) + \tausyneff \sum_k J_{jk} s_k(t-\delayeff) \,.
\end{align}
where $R$ is the resistance of the membrane, $\taumemeff$ the membrane time constant, $\tausyneff$ the synaptic time constant and $J_{jk}$ the synaptic weight of the connection from neuron $k$ to $j$.
Here the superscript $^\text{eff}$ indicates that these are not the parameters realized on hardware, but of an abstract model that we fit to hardware measurements.
Data were obtained from $100$ network realizations of the \rand{} case (same dataset as for \prettyref{fig:corr_calib}).
For each neuron we computed the STA of its free membrane potential individually for each spike train of its presynaptic partners.
The PSP for the current based model described above is given by
\begin{align}
  \label{eq:app_psp}
  V(t) =& \vmaxeff \frac{\tausyneff}{\tausyneff - \taumemeff} \left( e^{-(t-\delayeff)/\tausyneff} - e^{-(t-\delayeff)/\taumemeff} \right) \Theta(t-\delayeff) \,,
\end{align}
where we introduced $\vmaxeff = R J$ to simplify the fitting procedure.
For each pair of neurons $i$, $j$ we obtain a set of parameters: $\vmaxeff$, $\tausyneff$, $\taumemeff$ and $\delayeff$.

\begin{figure*}
\centering
\includegraphics{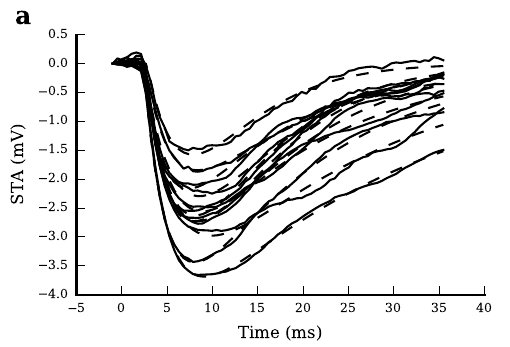}
\includegraphics{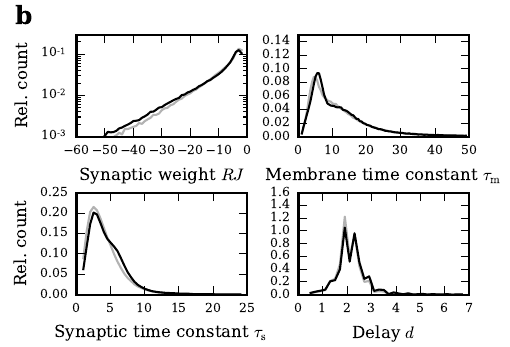}
\caption{
  \textbf{(a)} Fits of theoretical PSPs (dashed; for model, see \prettyref{eq:app_psp}) to STAs of free membrane potentials in the \fb{} case (solid).
  \textbf{(b)} Distributions of effective parameters obtained from fitting.
  Black and gray lines correspond to the fully calibrated ($a=0$) and uncalibrated ($a=1$) network, respectively.
  Median, 25th and 75th percentile of all fitted parameters are reported in \prettyref{tab:nordlie_params}.
  \label{appfig:sta_parameters}
}
\end{figure*}

The fitted curves (using the function {\it scipy.optimize.curve\_fit} with default parameters from Python's {\it scipy} module v0.17.0) match fairly well to the empirical data to allow an estimation of the {\it effective} values of the neuron parameters governing the shape of the postsynaptic potentials (\prettyref{appfig:sta_parameters}a).
However, note that the resulting parameter distributions do not precisely reflect the actual parameter distributions on the chip.

The distributions of neuron and synapse parameters change slightly with calibration.
For the calibrated network the absolute values of $\vmaxeff$, $\tausyneff$, $\taumemeff$ and $\delayeff$ are larger than for the uncalibrated network (see \prettyref{tab:nordlie_params} and \prettyref{appfig:sta_parameters}b).

\begingroup
\squeezetable
\begin{table*}

\setlength{\columnwidthleft}{0.2\fullfigwidth}
\setlength{\columnwidthmiddle}{0.2\fullfigwidth}

\begin{tabularx}{\fullfigwidth}{|p{\columnwidthleft}|X|}
  \hline\modelhdr{2}{A}{Model summary}\\\hline
  Populations    & One (inhibitory) \\
  \hline
  Topology       & - \\
  \hline
  Connectivity   & Random convergent connections (fixed in-degree) \\
  \hline
  Neuron model   & Leaky integrate-and-fire (LIF), fixed firing threshold, fixed absolute refractory time \\
  \hline
  Channel models & - \\
  \hline
  Synapse model  & Exponentially decaying currents, fixed delays \\
  \hline
  Plasticity     & - \\
  \hline
  External input & Resting potential higher than threshold (= constant current) ($\vrest>\vthresh$) \\
  \hline
  Measurements   & Spikes and membrane potentials \\
  \hline
  Other          & No autapses, no multapses \\
  \hline
\end{tabularx} \\

\begin{tabularx}{\fullfigwidth}{|p{\columnwidthleft}|p{\columnwidthmiddle}|X|}
  \hline\modelhdr{3}{B}{Populations}\\\hline
  \bf Name & \bf Elements & \bf Size \\
  \hline
  I & LIF neuron & $\netsize$ \\
  \hline
\end{tabularx} \\

\setlength{\columnwidthleft}{0.08\fullfigwidth}
\setlength{\columnwidthmiddle}{0.08\fullfigwidth}

\begin{tabularx}{\fullfigwidth}{|p{\columnwidthleft}|p{\columnwidthmiddle}|X|}
  \hline\modelhdr{3}{C}{Connectivity}\\\hline
  \bf Source & \bf Target & \bf Pattern \\
  \hline
  I & I & Random convergent connect, in-degree $\noinputs$ \\
  \hline
\end{tabularx} \\

\setlength{\columnwidthleft}{0.2\fullfigwidth}
\setlength{\columnwidthmiddle}{0.2\fullfigwidth}

\begin{tabularx}{\fullfigwidth}{|p{\columnwidthleft}|X|}
  \hline\modelhdr{2}{D}{Neuron and synapse model}\\\hline
  Type & Leaky integrate-and-fire, exponential currents \\
  \hline
  Subthreshold dynamics &
    Subthreshold dynamics ($t \not\in (t^*, t^* + \tauref)$): \newline
    \hspace*{1em} $\cmem \frac{\text{d}}{\text{d}t} \vm(t) = -\gleak (\vm(t) - \vrest) + \Isyn(t)$ \newline
    Reset and refractoriness ($t \in (t^*, t^* + \tauref)$): \newline
    \hspace*{1em} $\vm(t) = \vreset$ \\
  \hline
  Current dynamics &
    % For each presynaptic spike at time $t^*$ ($t > t^* + d$): \newline
    \hspace*{1em} $\tausyn \frac{\text{d}}{\text{d}t} \Isyn(t) = -\Isyn(t) + \sum_{i,k} J \delta(t-t_i^k)$ \newline
    Here the sum over $i$ runs over all presynaptic neurons and the sum over $k$ over all spike times of the respective neuron $i$ \\
  \hline
  Spiking &
    If $\vm(t^*-) < \vthresh \wedge \vm(t^*+) \ge \vthresh$: \newline
    \hspace*{1em} emit spike with time stamp $t^*$ \\
  \hline
\end{tabularx}

\begin{tabularx}{\fullfigwidth}{|p{\columnwidthleft}|X|}
  \hline\modelhdr{2}{E}{Measurements}\\\hline
  Spike trains & recorded from $\numrecordspikes$ neurons \\
  Membrane potentials & recorded from $\numrecordmem$ neurons \\
  \hline
\end{tabularx}

\caption{
  Description of the network model (according to \citep{Nordlie-2009_e1000456}).
  \label{apptab:sim_nordlie}
}

\end{table*}
\endgroup
\begingroup
\squeezetable
\begin{table*}

\setlength{\columnwidthleft}{0.15\fullfigwidth}
\setlength{\columnwidthmiddle}{0.3\fullfigwidth}

\begin{tabularx}{\textwidth}{|p{\columnwidthleft}|p{\columnwidthmiddle}|X|}
  \hline\parameterhdr{3}{B}{Populations}\\\hline
  \bf Name & \bf Values & \bf Description\\\hline
  $\netsize$  & $\{192,4800\}$ & network size \\
  \hline\parameterhdr{3}{C}{Connectivity}\\\hline
  \bf Name & \bf Values & \bf Description\\\hline
  $\noinputs$ & $\{15, 375\}$                  & number of presynaptic partners \\

  \hline\parameterhdr{3}{D}{Neuron}\\\hline
  \bf Name & \bf Values & \bf Description\\\hline
  $\cmem$    & \SI{0.2}{\nano\farad}           & membrane capacitance \\
  $\tauref$  & \SI{0.1}{\milli\second}         & refractory period \\
  $\vreset$  & \SI{-80}{\milli\volt}           & reset potential \\
  $\vrest$   & \SI{-52}{\milli\volt}           & resting potential \\
  $\vthresh$ & $\sim \mathcal{N}\left(-62, \lbrack 0, 8.8 \rbrack\right)$ \SI{}{\milli\volt}    & firing threshold \\
  $\gleak$   & \SI{10}{\nano\siemens}          & leak conductance \\

  \hline\parameterhdr{3}{D}{Synapse}\\\hline
  \bf Name & \bf Values & \bf Description\\\hline
  $\tausyn$   & \SI{5}{\milli\second}             & synaptic time constant \\
  $J$         & \SI{-0.254}{\nano\ampere}              & synaptic weight \\
  $\delay$    & \SI{1.}{\milli\second}            & synaptic delay \\
  %

  %\hline\parameterhdr{3}{D}{Input}\\\hline
  %\bf Name & \bf Values & \bf Description\\\hline
  %& & $\vrest > \vthresh$ \\
  %%
  %\hline

  %\hline\parameterhdr{3}{Other}{Emulation}\\\hline
  %\bf Name & \bf Values & \bf Description\\\hline
  %%
  %$\runtime$  & \SI{5}{\second} for calibration, \SI{10}{\second} else & emulation time in biological time domain \\
  %%
  %\hline

  \hline\parameterhdr{3}{E}{Measurements}\\\hline
  \bf Name & \bf Values & \bf Description\\\hline
  $\numrecordspikes$ & $\{192,4800\}$ & number of neurons spike trains are recorded from \\
  $\numrecordmem$ & $\{150,150\}$ & number of neurons membrane potentials are recorded from \\
  \hline
\end{tabularx}

\caption{
  Parameter values for the network model described in \prettyref{apptab:sim_nordlie} with distributed thresholds.
  \label{apptab:sim_nordlie_params}
}

\end{table*}
\endgroup
\begingroup
\squeezetable
\begin{table*}

\setlength{\columnwidthleft}{0.15\fullfigwidth}
\setlength{\columnwidthmiddle}{0.3\fullfigwidth}

\begin{tabularx}{\textwidth}{|p{\columnwidthleft}|p{\columnwidthmiddle}|X|}
  \hline\parameterhdr{3}{B}{Populations}\\\hline
  \bf Name & \bf Values & \bf Description\\\hline
  $\netsize$  & $192$ & network size \\
  \hline\parameterhdr{3}{C}{Connectivity}\\\hline
  \bf Name & \bf Values & \bf Description\\\hline
  $\noinputs$ & $15$                  & number of presynaptic partners \\

  \hline\parameterhdr{3}{D}{Neuron}\\\hline
  \bf Name & \bf Values & \bf Description\\\hline
  $\cmem$    & \SI{0.2}{\nano\farad}           & membrane capacitance \\
  $\tauref$  & \SI{0.1}{\milli\second}         & refractory period \\
  $\vreset$  & \SI{-80}{\milli\volt}           & reset potential \\
  $\vrest$   & \SI{-52}{\milli\volt}           & resting potential \\
  $\vthresh$ & \SI{-62}{\milli\volt}           & firing threshold \\
  $\gleak$   & \SI{10}{\nano\siemens}          & leak conductance \\

  \hline\parameterhdr{3}{D}{Synapse}\\\hline
  \bf Name & \bf Values & \bf Description\\\hline
  $\tausyn$   & \SI{5}{\milli\second}             & synaptic time constant \\
  % $J$         & \SI{0.254}{\nano\ampere}          & synaptic weight \\
  $J$         & $\sim\left[\mathcal{N}\left(0.254, \lbrack 0.00254, 0.75 \rbrack \right)\right]_+$ & synaptic weight, clipped to positive values \\
  $\delay$    & \SI{1.}{\milli\second}            & synaptic delay \\
  %

  %\hline\parameterhdr{3}{D}{Input}\\\hline
  %\bf Name & \bf Values & \bf Description\\\hline
  %& & $\vrest > \vthresh$ \\
  %%
  %\hline

  %\hline\parameterhdr{3}{Other}{Emulation}\\\hline
  %\bf Name & \bf Values & \bf Description\\\hline
  %%
  %$\runtime$  & \SI{5}{\second} for calibration, \SI{10}{\second} else & emulation time in biological time domain \\
  %%
  %\hline

  %
  \hline\parameterhdr{3}{E}{Measurements}\\\hline
  \bf Name & \bf Values & \bf Description\\\hline
  $\numrecordspikes$ & 192 & number of neurons spike trains are recorded from \\
  $\numrecordmem$ & 150 & number of neurons membrane potentials are recorded from \\
  \hline
\end{tabularx}

\caption{
  Parameter values for the network model described in \prettyref{apptab:sim_nordlie} with distributed weights.
  \label{apptab:sim_nordlie_params_weights}
}

\end{table*}
\endgroup

\begingroup
\squeezetable
\begin{table*}

\setlength{\columnwidthleft}{0.2\fullfigwidth}
\setlength{\columnwidthmiddle}{0.2\fullfigwidth}

\begin{tabularx}{\fullfigwidth}{|p{\columnwidthleft}|X|}
  \hline\modelhdr{2}{A}{Model summary}\\\hline
  Populations    & Two (excitatory, inhibitory) \\
  \hline
  Topology       & - \\
  \hline
  Connectivity   & Random convergent connections (fixed in-degree) \\
  \hline
  Neuron model   & Leaky integrate-and-fire (LIF), fixed firing threshold, fixed absolute refractory time \\
  \hline
  Channel models & - \\
  \hline
  Synapse model  & Exponentially decaying currents, fixed delays \\
  \hline
  Plasticity     & - \\
  \hline
  External input & Resting potential higher than threshold (= constant current) ($\vrest>\vthresh$) \\
  \hline
  Measurements   & Spikes and membrane potentials \\
  \hline
  Other          & No autapses, no multapses \\
  \hline
\end{tabularx} \\

\begin{tabularx}{\fullfigwidth}{|p{\columnwidthleft}|p{\columnwidthmiddle}|X|}
  \hline\modelhdr{3}{B}{Populations}\\\hline
  \bf Name & \bf Elements & \bf Size \\
  \hline
  E & LIF neuron & $\netsizeE$ \\
  I & LIF neuron & $\netsizeI$ \\
  \hline
\end{tabularx} \\

\setlength{\columnwidthleft}{0.08\fullfigwidth}
\setlength{\columnwidthmiddle}{0.08\fullfigwidth}

\begin{tabularx}{\fullfigwidth}{|p{\columnwidthleft}|p{\columnwidthmiddle}|X|}
  \hline\modelhdr{3}{C}{Connectivity}\\\hline
  \bf Source & \bf Target & \bf Pattern \\
  \hline
  E & E & Random convergent connect, in-degree $\noinputsE$, weight $\weightE$ \\
  E & I & Random convergent connect, in-degree $\noinputsE$, weight $\weightE$ \\
  I & E & Random convergent connect, in-degree $\noinputsI$, weight $\weightI$ \\
  I & I & Random convergent connect, in-degree $\noinputsI$, weight $\weightI$ \\
  \hline
\end{tabularx} \\

\setlength{\columnwidthleft}{0.2\fullfigwidth}
\setlength{\columnwidthmiddle}{0.2\fullfigwidth}

\begin{tabularx}{\fullfigwidth}{|p{\columnwidthleft}|X|}
  \hline\modelhdr{2}{D}{Neuron and synapse model}\\\hline
  Type & Leaky integrate-and-fire, exponential currents \\
  \hline
  Subthreshold dynamics &
    Subthreshold dynamics ($t \not\in (t^*, t^* + \tauref)$): \newline
    \hspace*{1em} $\cmem \frac{\text{d}}{\text{d}t} \vm(t) = -\gleak (\vm(t) - \vrest) + \Isyn(t)$ \newline
    Reset and refractoriness ($t \in (t^*, t^* + \tauref)$): \newline
    \hspace*{1em} $\vm(t) = \vreset$ \\
  \hline
  Current dynamics &
    % For each presynaptic spike at time $t^*$ ($t > t^* + d$): \newline
    \hspace*{1em} $\tausyn \frac{\text{d}}{\text{d}t} \Isyn(t) = -\Isyn(t) + \sum_{i,k} J \delta(t-t_i^k)$ \newline
    Here the sum over $i$ runs over all presynaptic neurons and the sum over $k$ over all spike times of the respective neuron $i$ \\
  \hline
  Spiking &
    If $\vm(t^*-) < \vthresh \wedge \vm(t^*+) \ge \vthresh$: \newline
    \hspace*{1em} emit spike with time stamp $t^*$ \\
  \hline
\end{tabularx}

\begin{tabularx}{\fullfigwidth}{|p{\columnwidthleft}|X|}
  \hline\modelhdr{2}{E}{Measurements}\\\hline
  Spike trains & recorded from $\numrecordspikesE$ excitatory and $\numrecordspikesI$ inhibitory neurons \\
  Membrane potentials & recorded from $\numrecordmemE$ excitatory and $\numrecordmemI$ inhibitory neurons \\
  \hline
\end{tabularx}

\caption{
  Description of the network model consisting of an excitatory and an inhibitory population (according to \citep{Nordlie-2009_e1000456}).
  \label{apptab:sim_nordlie_EI}
}

\end{table*}
\endgroup
\begingroup
\squeezetable
\begin{table*}

\setlength{\columnwidthleft}{0.15\fullfigwidth}
\setlength{\columnwidthmiddle}{0.3\fullfigwidth}

\begin{tabularx}{\textwidth}{|p{\columnwidthleft}|p{\columnwidthmiddle}|X|}
  \hline\parameterhdr{3}{B}{Populations}\\\hline
  \bf Name & \bf Values & \bf Description\\\hline
  $\netsizeE$  & $96$ & size of the excitatory population \\
  $\netsizeI$  & $96$ & size of the inhibitory population \\
  \hline\parameterhdr{3}{C}{Connectivity}\\\hline
  \bf Name & \bf Values & \bf Description\\\hline
  $\noinputsE$ & $7$                  & number of excitatory presynaptic partners \\
  $\noinputsI$ & $8$                  & number of inhibitory presynaptic partners \\

  \hline\parameterhdr{3}{D}{Neuron}\\\hline
  \bf Name & \bf Values & \bf Description\\\hline
  $\cmem$    & \SI{0.2}{\nano\farad}           & membrane capacitance \\
  $\tauref$  & \SI{0.1}{\milli\second}         & refractory period \\
  $\vreset$  & \SI{-80}{\milli\volt}           & reset potential \\
  $\vrest$   & \SI{-52}{\milli\volt}           & resting potential \\
  $\vthresh$ & \SI{-62}{\milli\volt}           & firing threshold \\
  $\gleak$   & \SI{10}{\nano\siemens}          & leak conductance \\

  \hline\parameterhdr{3}{D}{Synapse}\\\hline
  \bf Name & \bf Values & \bf Description\\\hline
  $\tausyn$   & \SI{5}{\milli\second}             & synaptic time constant \\
  $\weightE$  & \SI{0.0635}{\nano\ampere}        & excitatory synaptic weight \\
  $\weightI$  & \SI{-0.254}{\nano\ampere}         & inhibitory synaptic weight \\
  $\delay$    & \SI{1.}{\milli\second}            & synaptic delay \\
  %

  %\hline\parameterhdr{3}{D}{Input}\\\hline
  %\bf Name & \bf Values & \bf Description\\\hline
  %& & $\vrest > \vthresh$ \\
  %%
  %\hline

  %\hline\parameterhdr{3}{Other}{Emulation}\\\hline
  %\bf Name & \bf Values & \bf Description\\\hline
  %%
  %$\runtime$  & \SI{5}{\second} for calibration, \SI{10}{\second} else & emulation time in biological time domain \\
  %%
  %\hline

  %
  \hline\parameterhdr{3}{E}{Measurements}\\\hline
  \bf Name & \bf Values & \bf Description\\\hline
  $\numrecordspikesE$ & 96 & number of excitatory neurons spike trains are recorded from \\
  $\numrecordspikesI$ & 96 & number of inhibitory neurons spike trains are recorded from \\
  $\numrecordmemE$ & 150 & number of excitatory neurons membrane potentials are recorded from \\
  $\numrecordmemI$ & 150 & number of inhibitory neurons membrane potentials are recorded from \\
  \hline
\end{tabularx}

\caption{
  Parameter values for the network model described in \prettyref{apptab:sim_nordlie_EI}.
  \label{apptab:sim_nordlie_params_EI}
}

\end{table*}
\endgroup

\ifarxivmode
\else
  % set path via BSTINPUTS and BIBINPUTS in bashrc
  \bibliographystyle{neuralcomput_natbib_unsrt}
  \bibliography{bib}

  \end{document}
\fi
\fi

\end{document}